\documentclass{ar2e}	

\usepackage{ulem,graphicx,color,float,url,amsmath,multirow}
\usepackage{epstopdf}
\usepackage{amsfonts}
\usepackage{amssymb}

\DeclareGraphicsExtensions{.eps}

\usepackage{lineno}

\begin{document}
\input epsf.tex    
\input epsf.def   

\renewcommand\baselinestretch{1}

\input psfig.sty

\jname{}
\jyear{}
\jvol{}
\ARinfo{}

\title{Long-Baseline Neutrino Experiments}

\date{\today}
\author{M. V. Diwan
\affiliation{Brookhaven National Laboratory, Upton, NY, USA}
V. Galymov
\affiliation{Institut de physique nucl\'{e}aire de Lyon, Villeurbanne, France}
X. Qian
\affiliation{Brookhaven National Laboratory, Upton, NY, USA}
A. Rubbia
\affiliation{ETH Zurich, Institute for Particle Physics, Zurich, Switzerland}}

\begin{abstract} 
We review long-baseline neutrino experiments in which 
neutrinos are detected after traversing  macroscopic distances. 
Over such distances neutrinos have been found to oscillate among 
flavor states.  
Experiments with solar, atmospheric, reactor, and 
accelerator neutrinos have 
resulted in a coherent picture of  neutrino masses and 
mixing of the three known flavor states. We will summarize the current
best knowledge of neutrino parameters and phenomenology with 
our focus on the evolution of the 
experimental technique.  We will proceed  from the first evidence produced by 
astrophysical neutrino sources to the current open questions and the 
 goals  of future research.  
\end{abstract}

\maketitle
\thispagestyle{plain}



\section{Introduction}  \label{sec:introduction}  


Neutrinos are electrically neutral spin-1/2 particles that are emitted in radioactive  decays 
of unstable nuclei and  subatomic particles. They were originally proposed by W. Pauli in 
1930~\cite{brown1978} as an explanation of the apparent non-conservation of energy in radioactivity 
and inconsistencies between the exclusion principle and nuclear models of that time. 
Enrico Fermi was responsible for  the modern name of the particle and the theory of beta 
decay which is the foundation for current understanding of weak interactions~\cite{fermi1934}.  
Soon after the detection of free neutrinos from radioactive products in  nuclear 
reactors~\cite{Reines1956,  Cowan1956},  neutrino properties and their fundamental role 
in shaping the universe became an important line of inquiry~\cite{winter2000, kolb-turner}. 

Important questions regarding neutrinos are: the number and types of neutrinos,  
the nature and strength of their interactions with  matter,  their masses, 
and the  origin of these masses. A series of ground breaking experimental observations and 
theoretical advances since the 1950s have made progress on these issues~\cite{Agashe:2014kda}.  
While  neutrino physics historically had been   synonymous with  understanding of weak interactions, 
the  observation of neutrino oscillations has  opened a 
new  portal to fundamental questions regarding
 the relationship between quarks and leptons 
and the origin of fermion masses~\cite{theory-white-paper}. 

After the detection of neutrinos from nuclear reactors, 
the hypothesis 
of neutrino-antineutrino mixing was made in analogy to neutral 
mesons~\cite{ponte57-1, ponte1} given that only one type of neutrino was known at that time. 
This hypothesis was naturally extended to  mixing of neutrino flavors~\cite{ponte67-2, ponte2, Maki},
when the neutrinos produced in pion decays from a proton accelerator were shown to be
distinct from neutrinos produced in beta decays~\cite{twonus-1}. 
Subsequently,  the mixing of fields
was  realized to be a natural consequence of gauge theories with spontaneous symmetry 
breaking. Both quarks and neutrinos 
have now been shown to experience  flavor mixing 
although possibly originating from different mechanisms since
neutrinos do not have electric charge.  
Whereas quarks and 
charged leptons are   Dirac fermions with oppositely charged anti-particles, 
neutrinos could be either Dirac particles with a conserved leptonic charge or 
neutral Majorana particles with no conserved leptonic charge.  

\subsection{Neutrino Oscillations}\label{sec:oscillation}
Under the simplest assumption of  two-flavor neutrino mixing characterized by a single angle $\theta$, $\nu_e$ and $\nu_\mu$ are superpositions 
of two mass eigenstates $\nu_1$ and $\nu_2$ with masses $m_1$ and $m_2$:
\begin{equation}
\left(\begin{array}{c}\nu_e \\ \nu_\mu \end{array} \right) =
\left(\begin{array}{cc} \cos{\theta} & \sin{\theta} \\ 
-\sin{\theta} & \cos{\theta} \end{array} \right) \cdot \left(\begin{array}{c} \nu_1 \\ \nu_2 \end{array}\right). 
\end{equation} 
The time evolution of the $\nu_\mu$ state in vacuum 
in terms of $E_1$ and $E_2$, the energies of the two mass eigenstates, is given by:
\begin{equation}
\label{eq:2} 
\nu_\mu(t) = -\sin{\theta} \cdot e^{-iE_1 \cdot t} \cdot \nu_1+ \cos{\theta} \cdot e^{-iE_2 \cdot t} \cdot \nu_2. 
\end{equation} 
The complete quantum-mechanical description of neutrino oscillations requires the 
wave-packet formalism~\cite{smirnov1}. Nevertheless,  the correct expressions for 
the oscillation probabilities are obtained from Eq.  \ref{eq:2} by assuming that the two 
states have the same momentum or energy~\cite{Commins:1983ns}.  
The appearance probability to detect $\nu_e$ given an initial $\nu_\mu$ 
state at a distance $L$ is: 
\begin{equation}\label{eq1}
P(\nu_\mu \to \nu_e) = \sin^2{2 \theta} \cdot \sin^2 \left( 1.27\cdot \Delta m^2 \cdot \frac{L}{E} \right).
\end{equation} 
Given that the total probability must be conserved, the 
probability of $\nu_\mu$ survival or disappearance is simply: 
$P(\nu_\mu \to \nu_\mu) = 1- P(\nu_\mu \to \nu_e)$.   
Here  $\Delta m^2  = m^2_2 - m^2_1$ is in units of eV$^2$,  and $L/E$ in 
 $\rm km/GeV$ or $\rm m/MeV$. In the case of two-neutrino mixing, 
 the same formula applies to oscillations of antineutrinos. 

\begin{figure}[H]
\begin{centering}
\hspace*{-0.5in}
\includegraphics[width=0.95\textwidth]{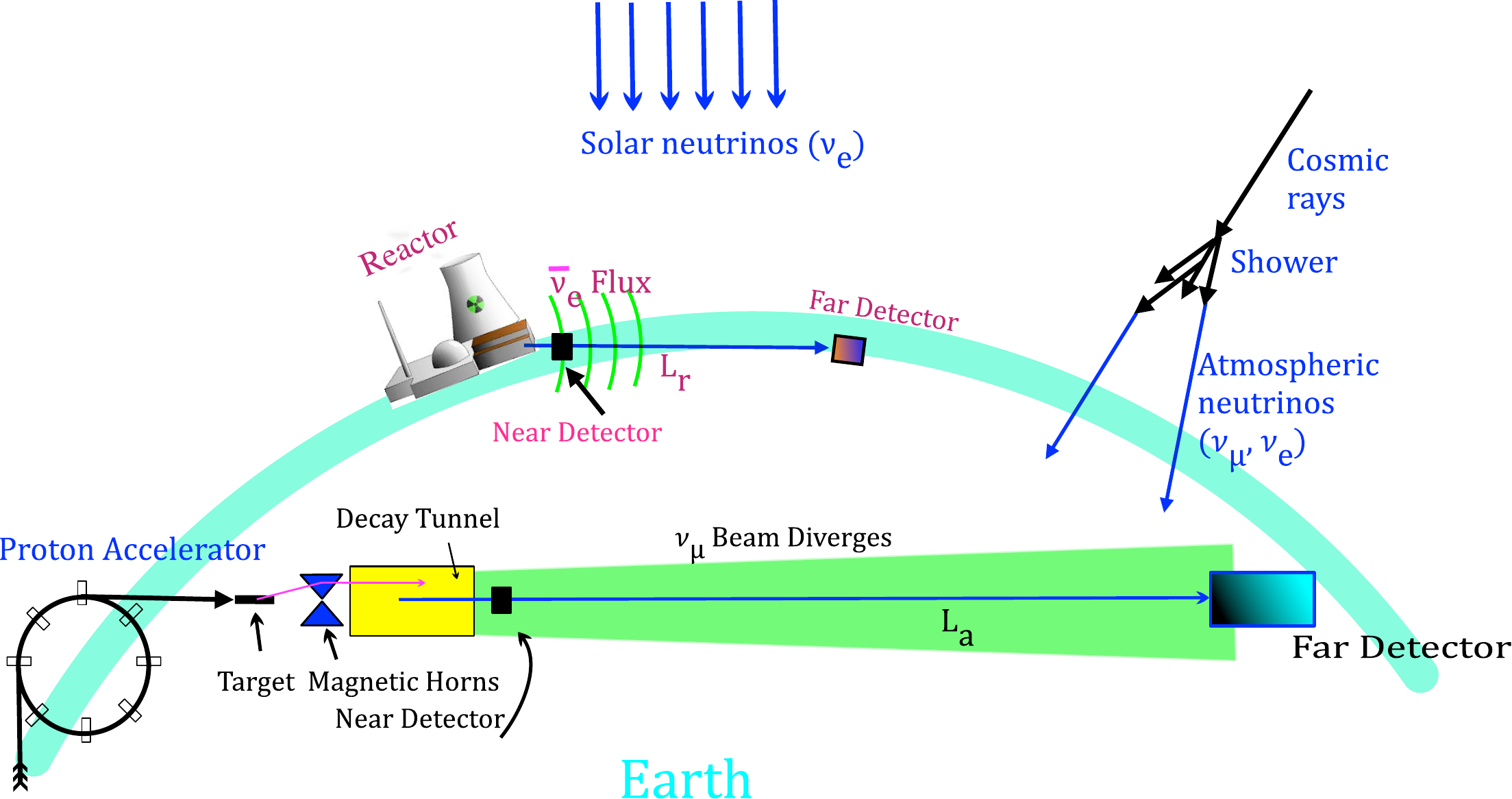}	
\par\end{centering}
\caption{\label{fig:lblart} 
Schematic illustrating neutrino sources  that have contributed 
to the current understanding of neutrino properties through neutrino oscillation experiments.  
Top: the Sun produces electron neutrinos ($\nu_e$).  Right:  neutrinos 
of two types,  $\nu_\mu$ and $\nu_e$,  and their antiparticles
are produced  by collisions of high energy cosmic rays with atoms in the Earth's atmosphere.  
Middle:   nuclear reactors emit electron antineutrinos ($\bar\nu_e$) isotropically.  
Bottom: high energy proton accelerators produce a beam of neutrinos, 
predominantly $\nu_\mu$ or $\bar\nu_\mu$  that is directed through the Earth.  
}
\end{figure}

When traveling through dense matter, neutrino  oscillation 
probabilities  can be significantly 
modified  by coherent forward scattering off electrons and 
nuclei\cite{Wolfenstein:1977ue,Mikheev:1986gs}. 
The equation of motion for neutrinos in the simple 2-flavor basis 
including  the effect of matter   is written as:
\begin{equation}
i \frac{d}{dt} \left( \begin{array}{c}  \nu_e \\ \nu_{\mu} \end{array} \right) =
{1\over 2}  \left( \begin{array}{cc} 
- ( \frac{\Delta m^2}{2 E}\cos 2\theta - \sqrt{2} G_F N_e )  &
 \frac{\Delta m^2}{2 E}  \sin 2\theta   \\
  \frac{\Delta m^2}{2 E}  \sin 2\theta  &   
( \frac{\Delta m^2}{2 E}\cos 2\theta - \sqrt{2} G_F N_e ) 
 \end{array} \right)
\left( \begin{array}{c}  \nu_e \\ \nu_{\mu}  \end{array} \right)  ~,
\label{eq:mat}
\end{equation}
 where $G_F$ is the Fermi constant, $N_e$ is the 
electron density which could be time (or position) dependent. 
The product  $\sqrt{2}G_F N_e$ acts as an extra potential 
due to  the  difference between $\nu_\mu$ and $\nu_e$ scattering amplitude  off electrons. 
This potential  reverses sign for antineutrinos.  
The eigenstates of this system  are not the vacuum mass eigenstates $\nu_1$ and $\nu_2$. 
Under the assumption of constant density  a solution in the form of Eq. \ref{eq1} can be obtained 
with the following substitutions: 
\begin{equation}\label{eq:mat2}  
\begin{array}{rcl}
\sin^2 2 \theta  & \to & \sin^22\theta^\prime =  \frac{\tan^2 2 \theta}{(1- {N_e\over N^{res}_e})^2 + \tan^2 2\theta} ~, \\ 
\Delta m^2 & \to & \Delta m^{\prime 2} = \Delta m^2 \sqrt{(1-{N_e\over N^{res}_e})^2 \cos^2 2 \theta + \sin^2 2 \theta} ~, \\
\end{array} 
\end{equation} 
where the quantity  
\begin{equation}\label{eq3}  
N^{res}_e = \frac{\Delta m^2 \cos 2 \theta}{2 E \sqrt{2} G_F} \approx 6.56 \times 10^6 {\Delta m^2 [{\rm eV^2}]\over E [{\rm MeV}]}  \cos 2 \theta \cdot N_A [{\rm cm^{-3}}]  
\end{equation} 
is called (for $\Delta m^2 \cos 2 \theta > 0$) the
 resonance density~\cite{Mikheev:1986gs,Barger:1980tf} with $N_A$ 
as Avogadro's number. 


Neutrino oscillations take place 
when the phases of  the eigenstates ($E_1 t$ and $E_2 t$) start 
to differ from each other as a function of time.  Over a  long period, 
even extremely small mass differences can lead to sufficient phase difference to cause 
flavor transformation.  
Observation of neutrino oscillations implies that neutrinos mix and at least one of 
them has mass. 
Neutrinos in the standard model of electroweak interactions  are  left-handed 
partners of the charged leptons and 
 only participate in  weak interactions.   The right-handed components of the neutrino field are  absent 
by definition making neutrinos massless in the model. 
Consequently, the individual lepton number is strictly conserved as confirmed 
in rare decays of muons, mesons, and collider experiments~\cite{Marciano:2008zz,Mihara:2013zna}.    
Inclusion of neutrino mass in the standard model by simply introducing right-handed partners 
creates  
the difficult problem of explaining the smallness 
and the nature of the neutrino mass  as either  Dirac or  Majorana. 
Observation of oscillations therefore is considered physics beyond
 the standard model that not only points to violation  of  
separate lepton number conservation  
but also forces us to consider that the nature of 
the neutrino mass~\cite{theory-white-paper} could be quite 
different from the other fermions.  
On the basis of observed oscillations and limits from direct measurements,  the electron neutrino mass
 is found to be at least five orders of magnitude 
smaller than the electron, the  lightest of the charged 
fermions~\cite{Otten:2008zz}.  But
 the existing data does not allow determination  of  neutrinos as  Dirac or Majorana fermions.

\subsection{Experimental Techniques}\label{sec:exp_tech}


Figure~\ref{fig:lblart} shows the natural and man-made sources of neutrinos 
that are the main contributors 
 to our understanding of neutrino oscillations. To be sensitive to oscillations
 the neutrino energy and the distance to the detector  need satisfy
 the condition $\Delta m^2 { L\over E}\gtrsim 1$ 
  for at least one  mass-squared difference. 
 In most cases, moving the detector or the source is impractical to cover
a range of L/E, and therefore a broad energy spectrum is beneficial to detect oscillations as a distortion 
of the expected energy distribution. 
For solar neutrino experiments ($L \simeq 10^{11}{\rm~m}$, $E \lesssim 15{\rm~MeV}$), 
for atmospheric neutrino experiments ($20 < L <  10000{\rm~km}$, $E\simeq  1{\rm~GeV}$),
for reactor neutrino experiments ($L \simeq 10 - 100000{\rm~m}$,  $E\simeq 3{\rm~MeV}$), and for accelerator 
neutrino experiments ($L \gtrsim 500{\rm~km}$, $E\simeq 1{\rm~GeV}$),  the  sensitivity ranges are
$\Delta m^2 \gtrsim 10^{-10},  10^{-4},   10^{-5},$ and $10^{-3} {\rm~eV^2}$, respectively.  
In all cases, the event rate per ton of detector  is very low 
because of the small cross section for neutrino interactions, necessitating  large detectors.  
The signal and background characteristics are technique dependent, 
nevertheless the detectors need to have 
very good shielding from cosmic ray muons generated in the atmosphere. 
This can be achieved by placing the 
detectors deep underground. 
In the case of accelerator neutrino beams, the pulsed nature of the beam 
provides additional background suppression.  

The Sun is a copious source of  $\nu_e$  with a 
flux of  $\sim 6 \times 10^{10}{\rm cm^{-2} sec^{-1}}$ at the Earth
produced by proton-proton and carbon-nitrogen-oxygen
 fusion reactions~\cite{Robertson:2012ib}. 
 The solar $\nu_e$ spectrum 
 is calculated using the standard solar
 model (SSM) with uncertainties ranging from $<$1\% for the $pp\to d e^+ \nu$ component 
 to $\sim$14\% for the higher energy Boron-8 component
 ($\lesssim 15 {\rm~MeV}$). 
 Solar neutrinos  are detected using 
charged-current neutrino reactions on nuclei with low thresholds or using elastic
 scattering from electrons. In the presence  of oscillations, 
a depletion of the expected $\nu_e$ flux is observed.

Nuclear fission reactors produce  electron antineutrinos 
generating $\sim2\times 10^{20} ~\bar\nu_e $ per GW of thermal power.
 As products of decay chains of neutron rich nuclei, 
these $\bar\nu_e$  are emitted isotropically from the
 core with an energy spectrum predominantly below $8 {\rm~MeV}$~\cite{Mueller:2011nm, Huber:2011wv}.   
They are detected via the inverse beta decay (IBD) reaction,  $\bar\nu_e + p \to e^+ +  n$,  
in detectors that contain free protons. 
The energy of an incoming antineutrino 
is determined by measuring the positron energy, while 
the detection of the recoiling neutron,  in coincidence with the 
positron signal,  suppresses backgrounds.  


Atmospheric neutrinos, coming from decays of $\pi$ and $K$ mesons 
produced in interactions of cosmic rays in the upper atmosphere,  
have an approximately isotropic flux of $\sim 4000 ~{\rm m^{-2} sec^{-1}}$ on the Earth. The dominant contributions come from the decay chains,
 $\pi^\pm  \to \mu^\pm  + \nu_\mu (\bar\nu_\mu)$ 
and $ \mu^\pm \to e^\pm + \nu_e  (\bar\nu_e) + \nu_\mu (\bar\nu_\mu)$. The flux has
a falling  energy spectrum  with a peak $\sim 1$ ~GeV,  and flux ratio
 $\Phi(\nu_\mu + \bar\nu_\mu)/\Phi(\nu_e+\bar\nu_e) \approx 2$~\cite{Gaisser:2002jj,Honda:2011nf}.  
Charged-current interactions of atmospheric neutrinos enable the study 
of oscillations over a broad range of distances ($\sim10$~km from above 
 to $1.3 \times 10^4$~km from below) in large underground detectors.
The sensitivity to oscillations is greatly improved by 
the well-understood  flux which
must be up-down symmetric (within $\lesssim 0.5\%$)  in the absence of oscillations \cite{Suzuki:2005id}.

Accelerator neutrino beams are produced by bombarding high energy protons into a
stationary target. 
The emerging $\pi$ and $K$ mesons 
are collected by a system of magnetic lenses (horns)~\cite{horns}, collimated
 towards the detector, and allowed to decay within a shielded  tunnel into muons and neutrinos.  
The  neutrino  energy can be in the range of $\lesssim 0.5 {\rm~GeV}$ to $\gtrsim 100 {\rm~GeV}$. 
The flux can be sign-selected to be predominantly
 $\nu_\mu$  or $\bar\nu_\mu$  with 
a small contamination ($\sim 1\%$)  from electron type.  
The primary proton energy, the geometry of the target, horns, as well as  the horn current
 can be adjusted to tune the beam spectrum with great flexibility.  

An important aspect of  reactor and accelerator 
experiments is the ability to place  near and  far detectors  to study  
the same source.
The near detector is placed sufficiently close to the source 
to measure the  neutrino flux before any significant flavor transformation. 
The far detector can be  positioned in the location of the
 expected oscillation maximum  to  measure the spectrum and flavor
 composition after oscillations.
 The comparison of observations at the two 
 sites allows for accurate determinations of oscillation  parameters 
free of many  uncertainties associated with
 neutrino production and interaction rates.     


 In Sec.~\ref{sec:pmns}, we 
review the 3-flavor neutrino model and summarize the experimental progress over  the past several
decades in establishing it. 
In Sec.~\ref{sec:current} and Sec.~\ref{sec:future}, we examine the current  and future planned 
long-baseline experiments  which  aim to determine the 
remaining unknowns and improve the
precision of  known neutrino parameters. We will describe the motivation for a  program of 
long-baseline neutrino experiments with particular emphasis on determination of 
the neutrino mass 
ordering and search for new charge-parity (CP) symmetry  violation. 

Although the discovery of neutrino oscillations took place using
 natural sources of neutrinos from the Sun 
and the atmosphere,  future precise  measurements are expected to
 originate from man-made sources such as reactors and 
accelerators that are well-monitored
and controllable.  As described in Sec. \ref{sec:future}, 
 future precise measurements of CP violation will only be possible in 
a long-baseline accelerator-based, high-power, and pure neutrino beam directed at 
 a near detector  and a very large 
and capable far detector.


\section{The Three-flavor neutrino Standard Model } \label{sec:pmns}

Although a definitive description of massive neutrinos within the
 standard model does not yet exist, all compelling neutrino oscillation data
can be described by mixing among three left-handed neutrino flavors, $\nu_e,
 \nu_\mu, \nu_\tau$ 
 in analogy with quark mixing via
the Cabibbo-Kobayshi-Maskawa (CKM) matrix~\cite{Agashe:2014kda}.  
This number of active neutrinos  is also compatible with the 
measured  invisible decay width of the Z-boson. 
The data firmly establishes that the  3 neutrino flavors  are superpositions of at least 3 light mass states with unequal masses, $m_1 \ne m_2 \ne m_3$, all smaller 
than $\sim1$ eV.  The oscillations  are characterized by two independent mass differences: 
$\Delta m^2_{21} = m^2_2 - m^2_1$ and $\Delta m^2_{32} = m^2_3 - m^2_2$ with $\Delta m^2_{31} = \Delta m^2_{32} + \Delta m^2_{21}$.    
The unitary $3\times 3$ mixing matrix, $U$,  
called the Pontecorvo-Maki-Nakagawa-Sakata (PMNS) matrix, 
is  parameterized by 3 Euler angles and depending on whether the 
$\nu_j$ are Dirac or Majorana, 1 or 3 phases
 potentially leading to CP violation. 
\begin{equation}\label{3nueq1}
\nu_{lL} = \sum_{j=1}^{3} U_{lj} \nu_{jL} 
\end{equation} 

\begin{eqnarray}\label{3nupmns} 
U =  \begin{pmatrix}
 c_{13}c_{12} & c_{13}s_{12} & s_{13} e^{-i \delta_{CP}}\\
  -c_{23}s_{12}-s_{13}s_{23}c_{12}e^{i\delta_{CP}} & 
c_{23}c_{12}-s_{13}s_{23}s_{12}e^{i\delta_{CP}} &
c_{13}s_{23} \\
 s_{23}s_{12}-s_{13}c_{23}c_{12}e^{i\delta_{CP}} & 
-s_{23}c_{12}-s_{13}c_{23}s_{12}e^{i\delta_{CP}} &
c_{13}c_{23} 
\end{pmatrix}   \nonumber   \\
\times diag(1, e^{i\frac{\alpha_{21}}{2}}, e^{i\frac{\alpha_{31}}{2}}) 
\end{eqnarray}
Here, $c_{ij}=\cos\theta_{ij}$ and $s_{ij}=\sin\theta_{ij}$, $\delta_{CP}$
 is the Dirac phase, and $\alpha_{21}$ and $\alpha_{31}$ are 
Majorana phases that cannot be observed in oscillation experiments. 
 Including the 3 neutrino masses 
 there are a total of 7 or 9 
additional free parameters in the minimally extended standard model 
with massive neutrinos
for the cases of Dirac or Majorana neutrinos, respectively. 
The full oscillation phenomenology is then described by 
modifying Eq.~\eqref{eq1}  for 3-$\nu$ mixing.  
For neutrinos produced with  energy $E$ and flavor $l$, the probability of its
 transformation to flavor $l'$ after 
traveling a distance $L$ in vacuum is expressed as:
\begin{eqnarray}\label{eq:osc_dis}
P_{ll^\prime} \equiv P(\nu_l\rightarrow \nu_{l'}) &  = &  \left |\sum_{i} U_{li}U^{*}_{l'i}e^{-i(m_{i}^2/2E)L} \right | ^2 \hfill \\ 
& = &  \sum_{i}|U_{li}U^*_{l'i}|^2 + \Re \sum_{i} \sum_{j \neq i} U_{li} U^{*}_{l'i} U^{*}_{lj} U_{l'j} e^{i\frac{\Delta m^2_{ij} L}{2E}}. \hfill \nonumber  
\end{eqnarray}
The  best values  for the parameters can be obtained from a global fit
to the data of neutrino oscillation experiments~\cite{Agashe:2014kda,Gonzalez-Garcia:2014bfa} 
The measured mass-squared  differences   are 
$\Delta m^2_{21} \approx 7.5 \times 10^{-5} {\rm~eV^2}$, 
 $| \Delta m^2_{32} | \approx 2.4 \times 10^{-3} {\rm~eV^2}$,
where the sign of $\Delta m^2_{32}$ is unknown and 
is commonly referred
 to as the problem of  mass ordering or hierarchy (MH).   
The determined mixing angles are
$\theta_{12} \cong 33.5^\circ$,  $\theta_{13} \cong 8.4^\circ$, 
and $\theta_{23} \approx \pi/4 + 3^o$ for normal 
 and $\theta_{23}  \approx \pi/4 -4.5^o$ for inverted  mass
ordering.  
Currently $\theta_{23}$ is the least known mixing angle with a strong degeneracy around $\pi/4$ 
because the measurement
is dominated by disappearance of $\nu_\mu$ which measures $\sin^2 2 \theta_{23}$. 
 From  global fits~\cite{Gonzalez-Garcia:2014bfa,Forero:2014bxa}
  to the data  the   current 1 $\sigma$  
uncertainties  on the parameters $\Delta m^2_{21}$, $\sin^2 \theta_{12}$, $|\Delta m^2_{32}|$, $\sin^2 \theta_{23}$, and $\sin^2 \theta_{13}$  
are  about 3\%, 4\%, 2\%, 12\%, and 5\%, respectively. 
Finally, the current experimental data does not significantly 
constrain the Dirac CP phase.

\begin{figure}[H]
\begin{centering}
\begin{tabular}{cc}
\vspace{-0.5cm} 
\includegraphics[width=0.95\textwidth]{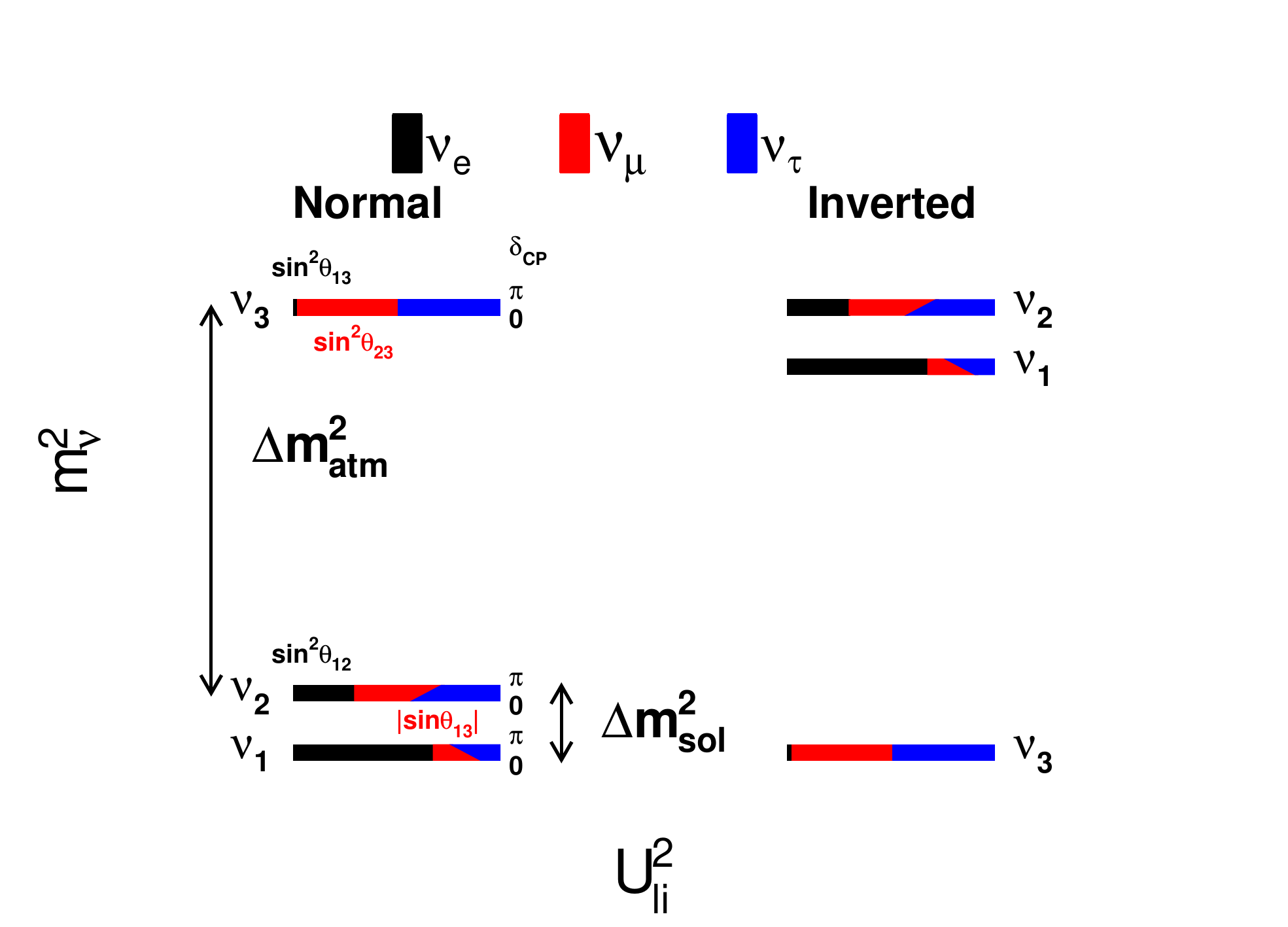}	
\end{tabular}
\par\end{centering}
\caption{\label{fig:mass}  
Patterns of the neutrino masses and mixing   for  normal (left) 
and inverted (right) ordering   \cite{Mena:2003ug}.  }
\end{figure}

Since  the   ratio of the two mass differences,
 $\Delta m^2_{32}/\Delta m^2_{21} 
\cong 33$,  is large,  most experimental situations can be analyzed using a 2-$\nu$ 
model with small corrections. If the experimental 
situation is such that $\Delta m^2_{21} L/(2E) \sim 1$ 
and $|\Delta m^2_{31}| L/(2E) \gg 1$ then oscillations due to the larger $\Delta m^2$ 
are averaged out due to either the size of the 
production region or experimental energy resolution. 
On the other hand,  if 
$|\Delta m^2_{31}| L/(2E) \sim 1$ then the 
oscillations due to the smaller 
$\Delta m^2_{21}$ remain a small correction. 
Fig.~\ref{fig:mass} illustrates the distinctive pattern of masses 
and mixings in the neutrino sector. 
The normal or NH  (inverted or IH)  mass ordering 
is  on the left (right).  
The maximal  $\nu_\mu/\nu_\tau$ mixing in $\nu_3$ is 
parameterized by $\theta_{23}$. The potentially large CP asymmetry
 is demonstrated as a variation of flavor content within each mass state.


\vspace{-0.5cm} 

\subsection{Measurement of $|\Delta m^2_{32}|$ and $\theta_{23}$  }

The first compelling evidence of neutrino flavor 
oscillation was obtained by the 
Super-Kamiokande (Super-K) experiment  observing  atmospheric neutrinos. 
 The cylindrical Super-K detector 
contains 50 ktons of pure water 
1000 m underground in  western Japan.  
Charged particles  with velocities exceeding that of light
 in water emit Cherenkov radiation which is detected using 
regularly spaced large photo-multiplier tubes along the walls,  sensitive 
to single photo-electrons~\cite{Fukuda:2002uc}. 
For atmospheric neutrinos, the detection principle relies on 
observing Cherenkov light from the charged lepton produced in the 
reaction: $\nu_{l} + N \rightarrow l + N^\prime$. 
 The energy, direction, and flavor of the incoming neutrino is 
 reconstructed from the detected pattern,  
arrival times, and intensity of Cherenkov photons.   

In 1998  Super-K reported a zenith angle 
dependent deficit of the upward going  atmospheric muon neutrinos,
inconsistent with the expected nearly isotropic  flux~\cite{Fukuda:1998mi}. 
This  definitive result
 followed the earlier hints from previous
water Cherenkov detectors,  Kamioka Nucleon Decay Experiment (KamiokaNDE) and Irvine-Michigan-Brookhaven (IMB) 
\cite{Hatakeyama:1998ea,Clark:1997cb}. 
The deficit of upward going 
muon neutrinos was not accompanied by excess of upward going electron neutrinos. 
This indicated that $\nu_\mu$  were oscillating into $\nu_\tau$ which 
 could not be directly observed in the detector due to the high energy threshold 
($\sim$3.5 GeV) required for  $\tau$ production 
 and the extremely short $\tau$ lifetime.   


The discovery of atmospheric neutrino
 oscillations was possible because of 
the large available dynamic range in $L/E$,  
the necessary event statistics provided by 
the massive underground detectors, 
and the favorably large $\theta_{23}$ mixing~\cite{kajita-annual-2014}.  
Super-K obtains the  best precision on the $\nu_{\mu} \rightarrow \nu_{\mu}$ 
parameters  through  selection of muon type events with good energy and angular resolution
to prepare an analysis using  $L/E$ as the observable. 
As demonstrated in \cite{Ashie:2004mr}  
the event depletion  at $\sim500 {\rm~km/GeV}$ 
corresponds to the first oscillation maximum 
with  
  $|\Delta m^2| \approx  (\pi/2) \times (1/500) {\rm~eV^2} $. 
At $L/E \gg 500 {\rm~km/GeV} $, the 
fast oscillations cannot be resolved, and an 
average factor of $1/2$ depletion of $\nu_\mu$ type events   is measured
corresponding to a precise measurement of $1-{1\over 2} \sin^2 2 \theta$
 or  near-maximal mixing.  
In the 3-neutrino scenario Super-K measures 
 $|\Delta m^2_{32}|$ 
(the atmospheric mass-squared splitting, $\Delta m^2_{atm}$) and a product of matrix elements
$4 |U_{\mu3}|^2\cdot (|U_{\mu1}|^2+|U_{\mu2}|^2)$. The
 latter further converts to a measurement 
of  $\theta_{23}$  (the atmospheric mixing angle). 

\begin{figure}[H]
\begin{centering}
\mbox{ \includegraphics[width=0.95\textwidth]{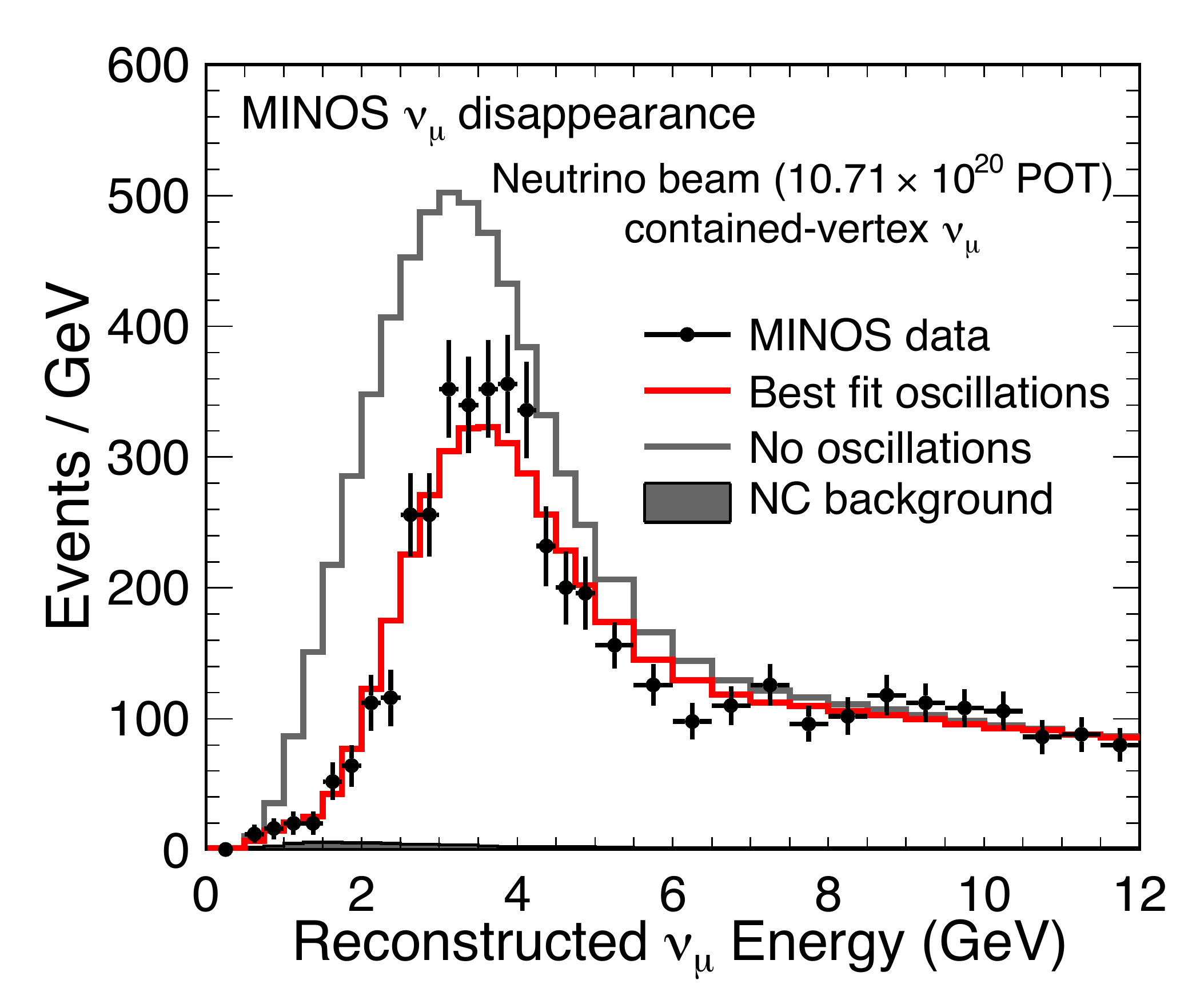}}
\par\end{centering}
\caption{\label{fig:mudisap} 
Distribution of muon neutrino events plotted as a function of reconstructed energy in 
the MINOS far detector~\cite{Adamson:2013whj} at a distance of 735 km.  
The solid grey line shows the expected spectrum under no 
oscillation hypothesis.  
The depletion of events at
$\gtrsim 1.5$ GeV   and the shape of the depleted spectrum  is consistent with dominant 
 $\nu_\mu \rightarrow \nu_\tau$ oscillations  with maximal mixing.  }
\end{figure}

The unusually large neutrino mixing observed in  the atmospheric oscillations 
discovery needed  a confirmation 
 by  accelerator neutrino experiments in which the neutrino beam
 is tunable and has high purity. 
Independent confirmation came from  
 K2K (KEK to Kamioka)~\cite{K2K}, MINOS(Main Injector Neutrino
Oscillations)~\cite{MINOS}, and T2K (Tokai to Kamioka)~\cite{Abe:2013fuq}
in which disappearance of muon 
neutrinos was observed with a laboratory produced beam.  
The most precise measurements of $|\Delta m^2_{32}|$ and $\theta_{23}$ 
are currently obtained by MINOS~\cite{Adamson:2014vgd} and T2K~\cite{Abe:2014ugx}, respectively. 
 Figure~\ref{fig:mudisap}  shows the muon neutrino spectrum  
observed in MINOS. 
MINOS uses an accelerator produced beam from the Fermilab
 (USA)  Main Injector accelerator with 120 GeV 
protons;  the peak of the low energy neutrino beam is adjusted to be about 3 GeV. 
The MINOS far detector is a magnetized steel 
scintillator detector with 5 ktons of mass located
 735 km away in the Soudan mine; in addition MINOS has a 
near detector using the same technology with 1 kton of mass
 placed $\sim$1 km away from the beam production target. A 
unique feature of MINOS is the ability to run the beam with 
changes to the target position and the horn current. 
By fitting several different spectra in the near detector, 
MINOS is able to make  predictions with $\lesssim 4\%$ uncertainty
~\cite{Adamson:2007gu}  for the  far detector event spectrum
 without oscillations which is important 
 for  parameter determination. 

T2K  uses an accelerator produced beam from the
 JPARC (Japan Proton Accelerator Research Complex) accelerator with 30 GeV protons; 
the Super-K water Cherenkov 
detector is used as the far detector located 295 km away. 
The direction of the beam is adjusted to be  
off-axis \cite{Beavis-BNL-52459}  by 2.5$^o$ 
to obtain a narrow band beam with maximum at 0.6 GeV \cite{Abe:2011ks} and  
 width of  $\sim 0.3$ GeV.  
The Super-K water Cherenkov detector has 
excellent energy and particle identification 
capability at this energy allowing the experiment
 to see almost complete disappearance of
 muon neutrino events at $L/E \sim 295/0.6~{\rm~km/GeV}$. 
The off-axis technique with excellent energy resolution 
 could provide  the best determination of  $\sin^2 2 \theta_{23}$.

In addition to confirming the $\nu_\mu$ disappearance signal, it was important to explicitly test for 
 $\nu_\mu \to \nu_\tau$ as indicated by the atmospheric neutrino result. Such a test requires 
detection of  $\tau$ leptons from a charged current interaction of $\nu_\tau$ at high energies. 
As the $\tau$ lepton decays almost immediately ($c \cdot \tau_{lifetime} \sim 87$ $\mu$ m), even a boosted 
$\tau$  with a momentum of $\sim$2 GeV/c would only travel  
about tenth of a mm before decay. Furthermore, the $\tau$ 
decays to a variety of final states making it 
difficult to distinguish from neutral current  events. 
Despite these difficulties  Super-K collaboration  has statistically
 identified  $\nu_\tau$ events at $3.8\sigma$ 
 in the atmospheric neutrino data from the backgrounds generated by deep
 inelastic scattering of $\nu_\mu$ and $\nu_e$ \cite{PhysRevLett.110.181802}.  
Detection of individual $\nu_\tau$ events requires 
tracking and detection of $\tau$ decays at very short distances. 
The  direct 
observation of $\nu_\tau$ appearance in a $\nu_\mu$ beam was
 performed by  the OPERA (Oscillation Project with Emulsion Tracking) 
experiment using 
 emulsion films with spatial resolution  of $\sim 1 ~\mu$m
\cite{Yoshida:2013pva}. 
OPERA has reported  observation of 5 $\tau$ 
events with an expected background of $0.25$ events
 confirming the dominance of the  
$\nu_\mu\rightarrow \nu_\tau$ oscillation mode  
\cite{Agafonova:2015jxn, PhysRevD.89.051102}.


\subsection{Measurement of $\Delta m^2_{21}$ and $\theta_{12}$}

Starting in 1967,  Ray Davis measured the solar $\nu_e$ flux  with 610 tons 
of liquid C$_2$Cl$_4$ through the reaction 
$\nu_e + ^{37}$Cl$\rightarrow e^- + ^{37}$Ar~\cite{Cleveland:1994er}.
The solar neutrino rate was determined by periodically counting decays of 
radioactive $^{37}$Ar extracted from the detector liquid. 
The measured $\nu_{e}$ flux was only about 1/3~\cite{Cleveland:1998nv}
 of the prediction from the 
standard solar model (SSM)~\cite{Bahcall:2000nu}. 
This result was further confirmed by the Gallium experiments,  
SAGE~\cite{Abdurashitov:1999zd} and GALLEX~\cite{Hampel:1998xg},  using the 
$\nu_e + ^{71}$Ga$\rightarrow^{71}$Ge$+e^-$  
reaction as well as   Kamiokande~\cite{Fukuda:1996sz}
and Super-K~\cite{Fukuda:1998fd,Fukuda:2002pe} experiments using the 
$\nu_e + e^- \rightarrow \nu_e + e^-$ reaction. 
Neutrino mixing  offered
a natural explanation to the solar neutrino  puzzle,
 since some of the $\nu_e$ generated in the Sun  could transform  to 
muon or tau neutrino flavors during flight  
and thus  become undetectable due to the low solar neutrino energies which 
are below the thresholds for  muon and tau production.

The Sudbury Neutrino Observatory (SNO)  experiment was designed to measure the flux of all 
neutrino flavors from the Sun using the neutral current 
reaction on deuterium, $\nu + d \rightarrow \nu + p + n$,
 and separately  the $\nu_e$ flux through   the 
charged-current reaction $\nu_{e} + d \rightarrow e^- +p + p$. 
SNO was a water Cherenkov detector with an active target 
mass composed of $\sim$ 1 kton of heavy water (D$_2$O) which provides 
the deuterium targets for solar neutrinos. The final-state neutron 
from the neutral-current reaction was detected 
by a variety of techniques with  consistent results. 
Measurement of the electron energy  from the final state of the 
charged current reaction
determines the solar neutrino spectrum above $\sim 5$ MeV 
\cite{Aharmim:2005gt}.   
The results~\cite{Ahmad:2002jz} are consistent with the
 predictions of SSM incorporating  $\nu$-oscillations 
 enhanced by the matter effect in the Sun.  
Borexino  in Italy  is  an extremely high resolution
 liquid scintillation underground detector  and has
 contributed by measuring the elastic 
scattering reaction,  $\nu  + e\to \nu + e$,  
 for $\nu_e$ energies below the $^7$Be line (0.861~MeV) in the solar neutrino spectrum. 
The  solar $\nu_e$ survival 
probability as  function of 
energy is obtained 
by combining data collected at all  energies by  the 
 radiochemical experiments(based on chlorine and gallium), Borexino, 
 SNO,  and Super-K  \cite{Bonventre:2013loa}.

The electron density in the core of the 
Sun is $N_e \sim 100 ~N_A  {\rm cm^{-3}}$ and
 it drops approximately exponentially 
to 0 at the surface.   An 
 electron neutrino of energy $\lesssim 1$ MeV   
generated in the core of the Sun has a resonance density 
$N^{res}_e$ much larger  than $N_e$ and therefore 
according to Eq.~\eqref{eq:mat2} the oscillations are as if in vacuum. 
The spread due to distance and the source size wash out the
 oscillations and the survival probability is a 
constant $\sim 1- \frac{1}{2}\sin^2 2 \theta_{12}$. 
 At energy $\gtrsim 5$ MeV, $N_{e}^{res} < N_e$ and 
 the $\nu_e$ eigenstate inside the Sun
coincides with the heavier effective mass
 eigenstate which adiabatically evolves
 into $\nu_2$ as the density decreases. 
 In this energy region the $\nu_e$ produced in the 
core leave the Sun in the $\nu_2$ state with 
the $\nu_e$ flavor content of 
$\sim \sin^2 \theta_{12}$.  
 This effect is called the Mikheyev-Smirnov-Wolfenstein (MSW)
effect~\cite{Wolfenstein:1977ue,MSW1,MSW2,MSW3}. 
 The energy dependent survival probability
additionally provides the constraint 
$\Delta m^2_{21} \cdot \cos 2 \theta_{12} > 0$ 
 and is a confirmation of the solar matter effect. 
 In the 3-$\nu$ picture, 
the charge current ($\nu_e$ only) and the 
neutral current (sum of all $\nu_{l}$) solar rate is 
a constraint on a combination of the PMNS matrix elements 
$|U_{e2}|^2\cdot(|U_{e1}|^2+|U_{e2}|^2) + |U_{e3}|^4$ or 
$\cos^{4}\theta_{13}\sin^2\theta_{12} + \sin^{4}\theta_{13}$. 

Depending on the solar zenith angle solar neutrinos pass through 
varying Earth matter density before they are detected.  The effect 
of Earth matter on solar neutrinos can produce a day/night asymmetry 
which was recently measured by Super-K~\cite{Renshaw:2013dzu} at 
2.7 sigma to be (-3.2$\pm$1.2 \%) consistent with expectations from 
the global fit.  The measurement provides an indication of the 
terrestrial matter effect as well as further independent constraint 
on the solar oscillation parameters.

\begin{figure}[H]
\begin{centering}
\mbox{ 
\includegraphics[width=0.95\textwidth]{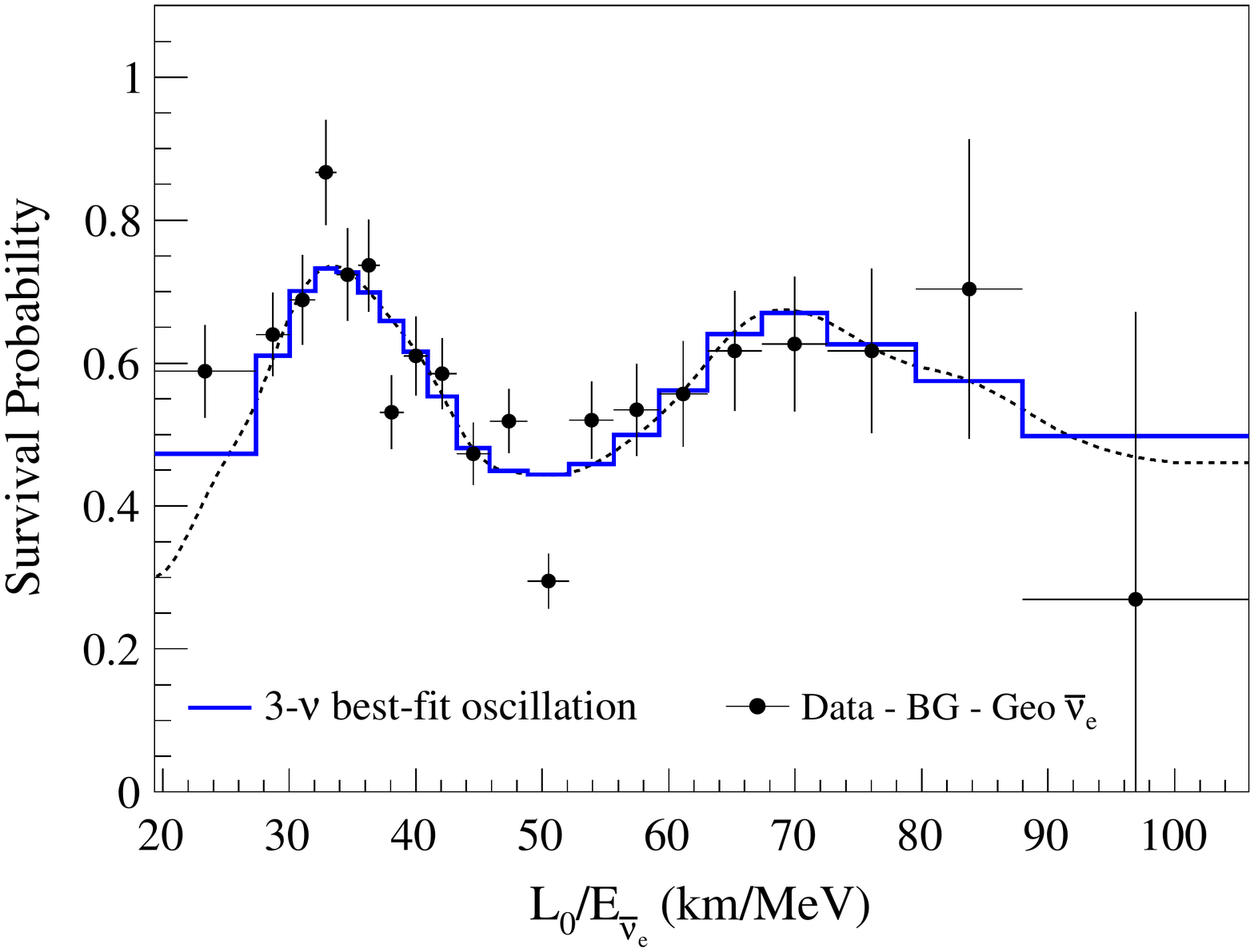}}
\par\end{centering}
\caption{\label{fig:edisap} 
Survival probability of $\bar\nu_e$ events as a 
function of $L/E$  in the KamLAND far detector~\cite{Gando:2013nba}.
The dip position of the oscillation 
($\sim$50 km/MeV) is consistent with the second oscillation node 
for  $\Delta m^2_{21}$. 
The size of the depletion is a measurement of  $\sin^22\theta_{12}$.}
\end{figure}

The Sun is a broadband source of pure $\nu_e$,  
 and when combined with the enhancement due to the 
MSW effect  and the very long distance to the 
Earth, it is not surprising that solar neutrino
 experiments provided the first hint of neutrino oscillations.  
Nevertheless,  despite consistency  
with the  large mixing angle $\theta_{12}$,
solar neutrino experiments included  
the possibility of multiple solutions at statistical significance of $\lesssim 3 \sigma$~\cite{Aharmim:2005gt} .  
The Kamioka liquid-scintillator antineutrino detector (KamLAND) experiment~\cite{Eguchi:2002dm} in
Japan selected the large mixing angle solution at $>5 \sigma$
 with detection of  reactor $\bar{\nu}_e$ 
over a long distance. There are a large number of  nuclear
power reactors around the KamLAND site with a flux averaged 
distance of $\sim$180 km. 
With neutrino energy determined from the measured positron energy in the IBD process, KamLAND was able to observe the $L/E$ oscillation spectrum shown in Fig. ~\ref{fig:edisap}.   
Taking advantage of the  very different $L/E$ ranges 
of the atmospheric and reactor  neutrinos, KamLAND and the solar neutrino 
experiments can be understood  in the 2-$\nu$  framework and provide  
 $\Delta m^2_{21}\sim7.5\times10^{-5}{\rm~eV^2}$
 (the solar mass-squared splitting, $\Delta m^2_{sol}$) 
and $\theta_{12}\sim34^{\circ}$ (the solar mixing angle). 
Within  the 3-$\nu$  framework  the KamLAND  
$\bar\nu_{e}$ disappearance constrains  $\Delta m^2_{21}$ and a combination
of PMNS matrix elements 
$2|U_{e3}|^2\cdot(|U_{e1}|^2+|U_{e2}|^2) + 4|U_{e1}|^2\cdot|U_{e2}|^2$. 
The latter is mostly sensitive to the 
solar mixing angle $\theta_{12}$ and has a weak 
dependence on the mixing angle $\theta_{13}$.

\vspace{-0.7cm}

\subsection{Measurement of $\theta_{13}$} 

The key observation to complete the 3-$\nu$ picture 
is the determination of 
$\nu_e$ oscillations in the same $L/E$ 
range as indicated by  the atmospheric $\Delta m^2_{32}$.  
The main impact of this observation is an 
explicit demonstration that the $\nu_e$ state is 
composed of at least three mass eigenstates or that all elements in 
the top row of the matrix in Eq.~\eqref{3nupmns} are non-zero. 
Both the solar and KamLAND measurements have a weak dependence on
 $\theta_{13}$. A joint analysis of these data 
 provided an initial  hint for a non-zero
 value ~\cite{Fogli:2011qn}. 
Previous reactor experiments
 with baseline of $\sim1$ km~\cite{Chooz1, Chooz2, PaloVerde} and a 
single far detector provided upper limits $\sin^2 2 \theta_{13} \lesssim 0.12$ at 90\% C.L.. 

A new campaign of experiments was launched to determine
$\theta_{13}$ by either $\bar\nu_e$ disappearance with  reactors 
 or $\nu_\mu \to \nu_e$ ($\bar\nu_\mu \to \bar\nu_e$)
 appearance with accelerators. 
In the case of reactor experiments, the 
 maximum expected depletion was $\lesssim 10\%$, and so a carefully designed 
experiment with near and far detectors was needed.
 In the case of accelerators,  the $\nu_e$ signal was expected to have 
 low statistics with background from both neutral current events and
 the $\nu_e$ contamination in the $\nu_\mu$ beam,  and therefore 
a well-designed neutrino beam and a large detector 
with excellent electron  identification were needed.

\begin{figure}[H]
\begin{centering} 
\includegraphics[width=0.95\textwidth]{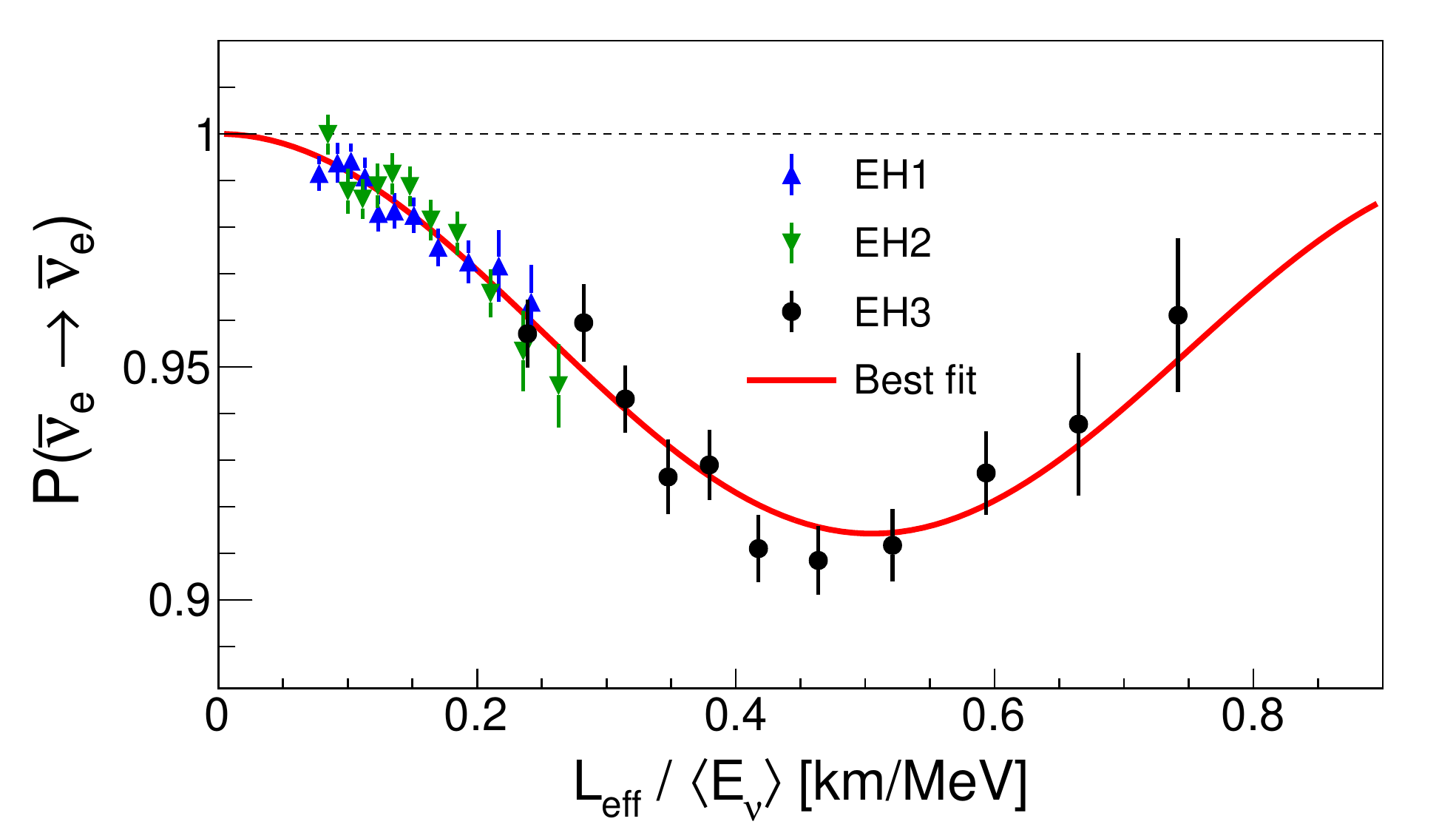}	
\par\end{centering}
\caption{\label{fig:theta13_global}  The measured $\bar{\nu}_e$ disappearance
probability is shown as a function of of $L/E$ from Daya Bay 
{\protect\cite{An:2015rpe}}. The associated oscillation frequency  corresponds to 
$\Delta m^2_{31} \sim \Delta m^2_{32} \sim 2.4 \times 10^{-3}{\rm~eV^2}$. 
} 
\end{figure}

In 2012, three new reactor experiments reported results on $\theta_{13}$.  Double Chooz (France)
reported results with a far detector only~\cite{dc} disfavoring
$\theta_{13}=0$  at 1.6$\sigma$. The 
Daya Bay experiment  (China) reported   the 
discovery of non-zero $\theta_{13}$ with  $\gtrsim 5 \sigma$ significance
from an array of three  near, and 
three  far identical detectors (each with 20 tons of gadolium-loaded liquid scintillator fiducial mass) 
 placed to optimize sensitivity from three nuclear power stations~\cite{dayabay}. 
RENO (Reactor Experiment for Neutrino Oscillations) 
also reported results from an array of nuclear power stations in Korea and 
identical near and far Gd-loaded 
liquid scintillator detectors  
confirming  Daya Bay's discovery with a $4.9\sigma$ significance~\cite{RENO}. 
 Daya Bay  increased their significance to $>$10$\sigma$ with a
larger data set~\cite{An:2013uza} and  recently 
added more target mass to the far and near sites (4 near and 4 far detectors)~\cite{An:2015rpe}.
The additional statistics and careful energy calibration in Daya Bay has resulted in 
  an independent measurement of  $\Delta m^2_{31}$ which governs this oscillation
(Fig.~\ref{fig:theta13_global}). 
The third neutrino mixing angle $\theta_{13}$ is now precisely measured to be 
 $\sim$8.4$^{\circ}$. 
MINOS~\cite{minos1} and T2K~\cite{t2k} have  reported their searches of 
$\nu_{\mu}$ to $\nu_e$ oscillation that is also sensitive to $\theta_{13}$. 
In particular, T2K obtained an early result  
disfavoring   $\theta_{13}=0$  at 2.5$\sigma$. 
The status of current accelerator experiments addressing $\nu_\mu \to \nu_e$ will be 
covered in more detail in the next section.




The picture of masses and mixings illustrated in Fig.~\ref{fig:mass} 
was assembled over several decades, mainly using
 precise  observations of $\nu_\mu$ or $\nu_e$  disappearance.   
By convention $m_2 > m_1$, and the strong
 evidence of matter effect in the Sun indicates 
 $\Delta m^2_{21} \cos 2\theta_{12} > 0 $ leading to 
 $\theta_{12} < \pi/4$~\cite{Agashe:2014kda}.     
But the question of neutrino mass hierarchy, whether 
$\nu_3$ is heavier or lighter than $\nu_{1,2}$, remains unresolved
~\cite{Qian:2015waa}. 
The sign of  $\theta_{23}-\pi/4$  
and the Dirac  or Majorana phases   are also unknown.
 These questions as well as further precision measurements 
 are expected to be 
addressed  by running and future  experiments 
optimized for the known parameters  
and focused on appearance measurements of $\nu_\mu \to \nu_e$  conversion.

\begin{figure}[H]
\begin{centering}
\hspace*{-0.5in}
\includegraphics[width=0.99\textwidth]{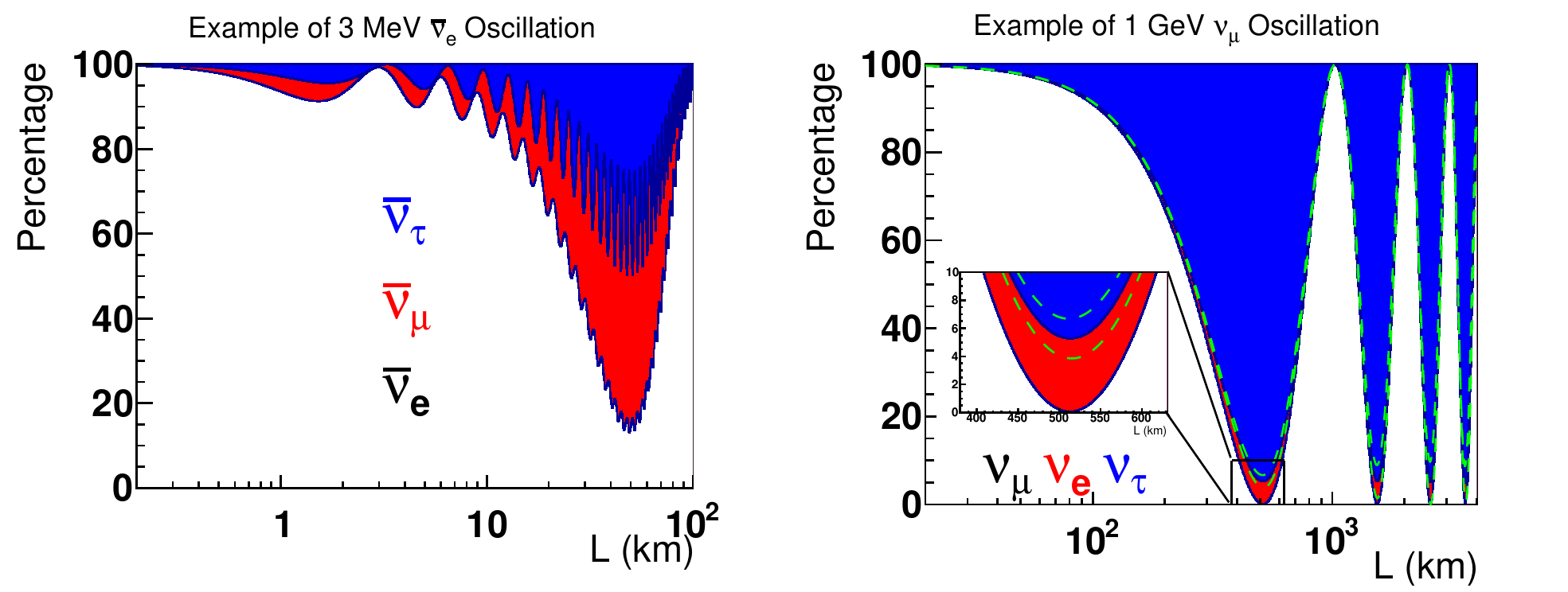}
\par\end{centering}
\caption{\label{fig:oscexample} 
Examples of neutrino oscillations in matter using the 3-neutrino oscillation 
model.  This calculation assumes $m_3 > m_2 > m_1$. 
The left hand plot displays the evolution of a 3 MeV electron  antineutrino.
The right hand plot displays the evolution of a 1 GeV muon type neutrino.  
}
\end{figure}

The precise prediction of 3 neutrino oscillations using the PMNS matrix model 
in matter is shown in Fig.~\ref{fig:oscexample}.  
For the left panel, the flavor evolution of a  3 MeV electron 
antineutrino  -- typical of a reactor experiment -- is shown 
over a distance of 100 km. 
Due to the large difference between $\Delta m^2_{21}$ and $|\Delta m^2_{32}|$,  
two different 
oscillation frequencies are seen with a relative shift dependent on the 
sign of $\Delta m^2_{32}$. 
 Reactor experiments can only observe $\bar{\nu}_e$ disappearance 
as the neutrino energies are too low for muon and tau production in charged-current reactions. 



For the right panel in Fig.~\ref{fig:oscexample}, 
the flavor evolution of  a 1 GeV muon neutrino  -- typical of an 
accelerator experiment -- 
is shown  over a distance 5000 km. 
According to the best fit parameters 
such  neutrinos 
should undergo a complete 
transformation after about $\sim$500 km with  a small fraction
appearing as $\nu_e$. 
The $\nu_e$ fraction 
undergoes a large modulation depending on the unknown CP phase (dashed lines represent 
$\delta_{CP}=\pm\pi/2$), and is opposite in magnitude for neutrinos and antineutrinos.  
The CP effect  also grows for higher oscillation nodes. 
A properly designed experimental program using reactor 
and accelerator sources should allow us to measure these oscillations over several nodes, 
to uncover if a large CP violation exists in the lepton sector, and to precisely 
determine   the parameters including the mass hierarchy and  the possibility  of  maximum mixing.

A few anomalous experimental results including  LSND~\cite{LSND},
 MiniBooNE~\cite{miniboone}, 
and measurements of absolute reactor flux~\cite{anom}  
 cannot be explained by the 3-neutrino  framework. 
These results  suggest that the 3$\times$3 PMNS matrix 
might  not be unitary,  and there might exist sterile neutrino(s) with 
the mass-squared splitting $\Delta m^2_{41} \sim 1 {\rm~eV^2} \gg
|\Delta m^2_{32}|\sim|\Delta m^2_{31}| \gg \Delta m^2_{21}$ with 
small mixing with the three active neutrinos.     
Details about these data and future 
direct searches at short baselines ($L/E \sim 1 $ km/GeV)
 can be found in Ref.~\cite{Abazajian:2012ys}. 
A joint fit of  the current solar, KamLAND, and Daya Bay results 
provides the first unitarity test of the top row of the 
PMNS matrix ($|U_{e1}|^2 + |U_{e2}|^2 + |U_{e3}|^2 - 1 \approx \pm 0.04$),
insufficient to exclude  sterile neutrinos with small mixing.    
~\cite{unitarity_decay, Qian:2013ora}.  
The comparison of $\theta_{13}$  derived from reactor 
disappearance measurements and from accelerator  appearance 
measurements  also provides a strong indirect test of the unitarity of PMNS 
matrix~\cite{Qian:2013ora}. 
In the rest of this review, we will focus on the standard three-flavor 
neutrino model and precise measurements of 
its parameters with current and future long-baseline experiments.

\section{Phenomenology of CP violation and mass hierarchy}

\label{sec:pheno}



We first examine  the appearance probability 
$P_{\mu e}$ in the 3-flavor model
to understand the experimental optimization:
\begin{equation}
P_{\mu e} = | U_{\mu1}U^*_{e_1}e^{-iE_1 t}  + U_{\mu2}U^*_{e_2}e^{-iE_2 t} + U_{\mu3}U^*_{e_3}e^{-iE_3 t}|^2
\end{equation}
where $E_i t$ represents the phase advance 
of the mass eigenstate $\nu_i$  over the flight time between
 the source and the detector.   
The oscillation probability for $\nu_e\rightarrow \nu_\mu$
or $\bar\nu_\mu\rightarrow \bar\nu_e$
 can be 
obtained by simply exchanging the labels $(e)\leftrightarrow(\mu)$. 
Therefore, if any element of the PMNS matrix is complex,  
(e.g. $U_{e3}\neq U^*_{e3}$),
 neutrino oscillations are not 
invariant under time reversal or  charge-parity conjugation. 
This is a phenomenological 
consequence of 3-generation mixing with at least 
one complex phase.  
%
%



Explicitly,  if the PMNS matrix is complex, then
 the asymmetry $a^{ll'}_{CP}\equiv P(\nu_l\to \nu_{l'}) -
 P(\bar\nu_l \to \bar\nu_{l'})$ is a direct measurement of the CP violation
 in the neutrino system. 
In the case of 3-neutrino mixing, $a^{\mu e}_{CP} = -a^{\tau e}_{CP} = a^{\tau \mu}_{CP}$.  
We prefer to use the fractional asymmetry  in experimental considerations:  
$A^{ll'} \equiv  ({P(\nu_l\to \nu_{l'}) -
 P(\bar\nu_l \to \bar\nu_{l'})})/({P(\nu_l\to \nu_{l'}) +
 P(\bar\nu_l \to \bar\nu_{l'})})$. The preferred observable is  $A^{\mu e}$
since it  is expected to be large, 
and the production and detection
 of $\nu_\mu$ and $\nu_e$ is far easier than $\nu_\tau$. 
For vacuum oscillations the CP asymmetry to leading order in 
$\alpha = \Delta m^2_{21}/\Delta m^2_{31}$ is: 
\begin{equation} 
A^{\mu e}_{CP} \approx  {2 J^{PMNS}_{CP}\over \sin^2(\theta_{23}) \sin^2(\theta_{13}) \cos^2(\theta_{13})
}  \times \alpha  
{\Delta m^2_{31} L \over {4 E_\nu}},
\end{equation} 
where $J^{PMNS}_{CP}$ is the rephasing invariant~\cite{Agashe:2014kda}, 
\begin{equation} 
\label{eq:jcp} 
 J_{CP}^{PMNS} \equiv \frac{1}{8} \sin 2 \theta_{12} \sin 2 \theta_{13}
\sin 2 \theta_{23} \cos \theta_{13} \sin \delta_{CP} \approx 0.032 \sin \delta_{CP}.  
\end{equation} 
This is in sharp contrast to the very 
small mixing in the quark sector which leads to the 
very small value for the corresponding 
invariant: $J^{CKM}_{CP} \approx 3\times 10^{-5}$, 
despite the large value of $\delta^{CKM}_{CP}\sim 70^o$. 
Recent  studies in leptogenesis \cite{Fukugita:1986hr} 
have shown that the phases in the PMNS matrix can provide the 
CP violation large enough  for the 
generation of observed baryon asymmetry of
 the universe \cite{Abada:2006fw, Nardi:2006fx}.
Therefore 
observation of  CP violation due to $J^{PMNS}_{CP}$ could
 account for 
a large  fraction of the  baryon asymmetry
\cite{Pascoli:2006ie, Pascoli:2006ci}.    
The asymmetry $A^{\mu e}_{CP}$  increases 
as a function of $L/E$ and is shown to decrease with $\sin\theta_{13}$ 
\cite{marciano:2001tz, Parke:2005ev} as long as it is not too small,  
and with current parameters the asymmetry can 
be as large as  $\sim 32\%$ for $L/E \sim 500 {\rm~km/GeV}$  
\cite{Bass:2013vcg}.     Due to the linear dependence on $L$, and
 $\sin\theta_{13}$, the statistical error on  $A^{\mu e}_{CP}$ is 
approximately independent of  $L$ and $\theta_{13}$ providing 
substantial flexibility in experimental 
considerations~\cite{marciano:2001tz,Diwan:2003bp}.

Observation of $A^{\mu e}_{CP}$  requires a 
 long-baseline experiment with a pure accelerator generated 
 beam of $\nu_\mu$ and $\bar\nu_\mu$  crossing the
Earth before reaching the detector.  The effect of Earth 
matter must be taken into account 
 using  the electron density through the  crust or mantle 
of  $N_e \cong  1.8 N_A {\rm~cm}^{-3}$.   
Using the  well-known values of $|\Delta m^2_{32}|$ and $\theta_{13}$, 
for typical accelerator neutrino energy of $\sim 1-10 {\rm~GeV}$
 large   change is expected to 
 the oscillation probability (Eq.~\eqref{eq:mat} and \eqref{eq3}).  
 The effect for neutrinos (anti-neutrinos) leads to an 
enhancement (suppression) 
 for $m_3 > m_2 > m_1$ and a suppression (enhancement) for $m_1 < m_2 < m_3$, 
and therefore measuring it 
will determine the order  of the neutrino masses.  
Atmospheric neutrinos are also expected to show 
significant sensitivity to Earth matter for $E_\nu > 2 {\rm ~GeV}$
  crossing the core \cite{Chizhov:1999he, Akhmedov:2006hb}. 
The  effect of Earth matter on neutrino oscillations 
with accelerator or atmospheric neutrinos 
has not been demonstrated in a definitive way, 
therefore such a measurement has interest both for  phenomenology and   
for determining the mass ordering. 


Since the size of the effect on $P_{\mu e}$ 
due to the CP phase and the Earth matter is 
similar, both effects have to be considered for actual experiments.  
For accelerator experiments,  assuming 
a constant density of matter,  a  sufficiently accurate expression 
to leading order in 
$\alpha = \Delta m^2_{21}/\Delta m^2_{31}$ 
 has been derived \cite{Freund:2001pn}(also see Fig. \ref{fig:oscexample}): 
\begin{eqnarray}
\label{eq:nueapp} 
P(\nu_\mu \rightarrow \nu_e)  &  = & \sin^2\theta_{23} \frac{\sin^22\theta_{13}}{(A-1)^2} \sin^2[(A-1)\Delta_{31}]  \hfill   \\
&+  &  \alpha^2\cos^2\theta_{23}\frac{\sin^22\theta_{12}}{A^2}\sin^2(A\Delta_{31})  \hfill \nonumber  \\
&- &   \alpha \frac{\sin2\theta_{12}\sin2\theta_{13}\sin2\theta_{23}\cos\theta_{13}\sin\delta_{CP}}{A(1-A)}\sin\Delta_{31}\sin(A\Delta_{31}) \sin[(1-A)\Delta_{31}]  \hfill \nonumber \\ 
&+  &  \alpha \frac{\sin2\theta_{12}\sin2\theta_{13}\sin2\theta_{23}\cos\theta_{13}\cos\delta_{CP}}{A(1-A)}\cos\Delta_{31}\sin(A\Delta_{31})\sin[(1-A)\Delta_{31}],  \hfill \nonumber 
\end{eqnarray}
where 
\begin{equation}
\Delta_{ij} = \Delta m^2_{ij}L/4E_{\nu}, ~A=\sqrt{2}G_{F}N_{e}2E_{\nu}/\Delta m^2_{31}. \nonumber
\end{equation}
For  anti-neutrinos the signs of  $\delta_{CP}$ and   $A$  are reversed. The last
 two terms in the expression are proportional  to $J^{PMNS}_{CP}$.  
The dependence on the mass 
hierarchy in $A$ and the  CP phase 
can be disentangled with precise measurement of 
 $\nu_\mu \to \nu_e$ and $\bar\nu_\mu \to \bar\nu_e$ appearance 
 as a  function of energy \cite{Diwan:2003bp,Diwan:2004bt}. 


The precise measurement of $\sin^{2}2\theta_{13}$ by  reactor neutrino 
experiments~\cite{An:2013zwz,2013arXiv1312.4111S}
has also provided a unique opportunity to determine the neutrino mass hierarchy 
in a medium-baseline high resolution  reactor neutrino oscillation
 experiment~\cite{Petcov:2001sy,Choubey:2003qx,deGouvea:2005hk,Learned:2006wy,Minakata:2007tn,Parke:2008cz,Zhan:2008id,Zhan:2009rs,Qian:2012xh,Ciuffoli:2012iz,Ge:2012wj,Li:2013zyd}.
The energy and distance for reactor experiments is 
sufficiently low that the vacuum 
 formula for the survival probability  of 
an electron antineutrino can be used (see Fig.~\ref{fig:oscexample}):
\begin{eqnarray}
\label{eq:medreact} 
P(\nu_e \to \nu_e) & = &  1   \hfill \\  
& - & \cos^4(\theta_{13}) \sin^2(2\theta_{12})\sin^2(\Delta_{21}) \hfill \nonumber   \\ 
& - & \cos^2(\theta_{12}) \sin^2(2\theta_{13})\sin^2(\Delta_{31}) \hfill  \nonumber \\ 
& - & \sin^2(\theta_{12}) \sin^2(2\theta_{13})\sin^2(\Delta_{32}), \hfill   \nonumber 
\end{eqnarray}
This probability depends on the mass hierarchy  as
 $|\Delta_{31}|> |\Delta_{32}|$ for  NH  and 
 $|\Delta_{31}|< |\Delta_{32}|$ for IH.
The measured reactor neutrino energy spectrum for $\Delta m^2_{21}L/(2E) \sim 1$
 (or $L\sim 50 {\rm km}$)  
  is expected to exhibit oscillations with 
 a slow ($\Delta_{21}$)  and a fast component.     
The fast component arising from the $\Delta_{31}$ and $\Delta_{32}$ terms 
has  amplitude proportional to $\sin^2 (2\theta_{13})$. 
The two choices for mass hierarchy produce  a 
 small energy dependent shift (of the order of $\Delta m^2_{21}/\Delta m^2_{32} \sim \pm 3\%$)
  in the oscillation pattern which can be measured with sufficient energy resolution.  
From above formula, well-optimized reactor experiments can access 
five neutrino mixing  parameters: $\theta_{12}$,
 $\Delta m^2_{21}$, $\theta_{13}$, $|\Delta m^2_{32}|$,
and the mass hierarchy (or the sign of $\Delta m^2_{32}$). Such 
high precision measurements will enable  tests of neutrino mass and 
mixing models. 
For example, models of 
 quark-lepton complementarity and others~\cite{Minakata:2004xt, Altarelli:2010gt},  
have predicted $\theta_{12} + \theta_{Cabbibo} = \pi/2$ and 
$\theta_{13}\approx \theta_{Cabbibo}/\sqrt{2}$. 


Based on  Eq.~\eqref{eq:nueapp} and Eq.~\eqref{eq:medreact}, 
an optimized program of measurements for 
appearance  for $\nu_\mu \rightarrow \nu_e$ using an accelerator $\nu_\mu$ beam, 
and disappearance for $\bar\nu_e\to \bar\nu_e$ using reactor flux cover the remaining issues within 
neutrino 
oscillation physics. They also provide redundancy
 for measurements of mass hierarchy, $\Delta m^2_{32}$,   
and $\theta_{13}$. These measurements depend on man-made pure beam sources
with optimized  $L/E$ range. 
 Neutrino physics phenomenology has proven to be quite rich
and any future program should not only be guided by  measurements motivated by the current 
best model, but  should also be sufficiently redundant and with enough dynamic range 
 to allow identification  of new physics effects if they exist.  
In the next sections we will examine the current and future planned programs for 
these measurements.

\section{Current-generation of Experiments }
\label{sec:current} 

The accelerator disappearance experiment MINOS+ \cite{Timmons:2015laq}, 
and the reactor experiments Daya Bay, Double Chooz, and RENO
are expected to continue operations to improve  
 precision on $\Delta m^2_{32}$, $\sin^2 2 \theta_{23}$ and 
$\sin^2 2 \theta_{13}$.  
The error on  the value of the $\sin^2{\theta_{13}}$ 
is expected to reach {$\sim$3\%}~\cite{Zhang:2015fya}. 
The sign and magnitude of $\theta_{23}-\pi/4$   is 
of significance for the 
theoretical work aimed at explaining underlying symmetries responsible 
for  neutrino mixing \cite{Altarelli:2010gt}. 
The measurement of  $\delta_{CP}$ is correlated to  $\theta_{23}$ 
since, to leading order (Eq.~\eqref{eq:nueapp}), 
the appearance probability depends on $\sin^2{\theta_{23}}\cdot\sin^2{2\theta_{13}}$. 
Precision measurement of  $\theta_{23}$ is therefore an important 
aspect of the  current program. 
The accumulation of atmospheric and solar neutrino data is also  expected to 
continue with Super-K, MINOS+, IceCube~\cite{Aartsen:2014yll}, and Borexino. 



The main focus of the current program, however, is 
 the $\nu_\mu\rightarrow \nu_e$ transition with the T2K and NO$\nu$A 
experiments.  These
experiments were both optimized before the discovery of $\theta_{13}$ 
 to obtain evidence for this transition 
with the best signal to background ratio.  
The two principal backgrounds to an electron neutrino event 
in the few GeV range are: 
the intrinsic $\nu_e$ contamination present in the 
accelerator-produced  beam at $\sim 1\%$ level, 
and  weak neutral current $\nu$   interactions 
that  produce photons and $\pi^0$ particles.  
The $\nu_e$  background in the beam  comes
 from $\mu$ and $K$-meson decays in the decay tunnel. 
The $\nu_e$ 
background is indistinguishable from the signal except that it 
has a broader energy spectrum.  
The neutral current  processes 
produce photons from $\pi^0\to \gamma \gamma$ 
decays that could be misidentified as single electrons
in case of asymmetric or overlapping 
electro-magnetic showers. 
%
To limit the 
impact of both of these  backgrounds with broad energy spectra,  
both T2K and NO$\nu$A have adapted the strategy of using 
a  narrow neutrino energy spectrum. 
The narrow energy spectrum cuts down the 
contributions from neutrinos outside of the energies of 
interest as set by the oscillation probability. 
The narrow-band  beam is achieved with the off-axis neutrino beam 
technique \cite{Beavis-BNL-52459} in which the detector is placed 
at a small angle to the beam to exploit the kinematic momentum peak
in the 2-body   $\pi  \to \mu \nu_\mu$ decay.    
The T2K and NO$\nu$A experiments 
utilize  neutrino beams with a peak energy 
of $\sim$0.6 GeV and $\sim$2 GeV, at off-axis angles of 44 
and 14 mrad, respectively. 

\begin{figure}[H]
\begin{centering}
\includegraphics[width=0.95 \textwidth]{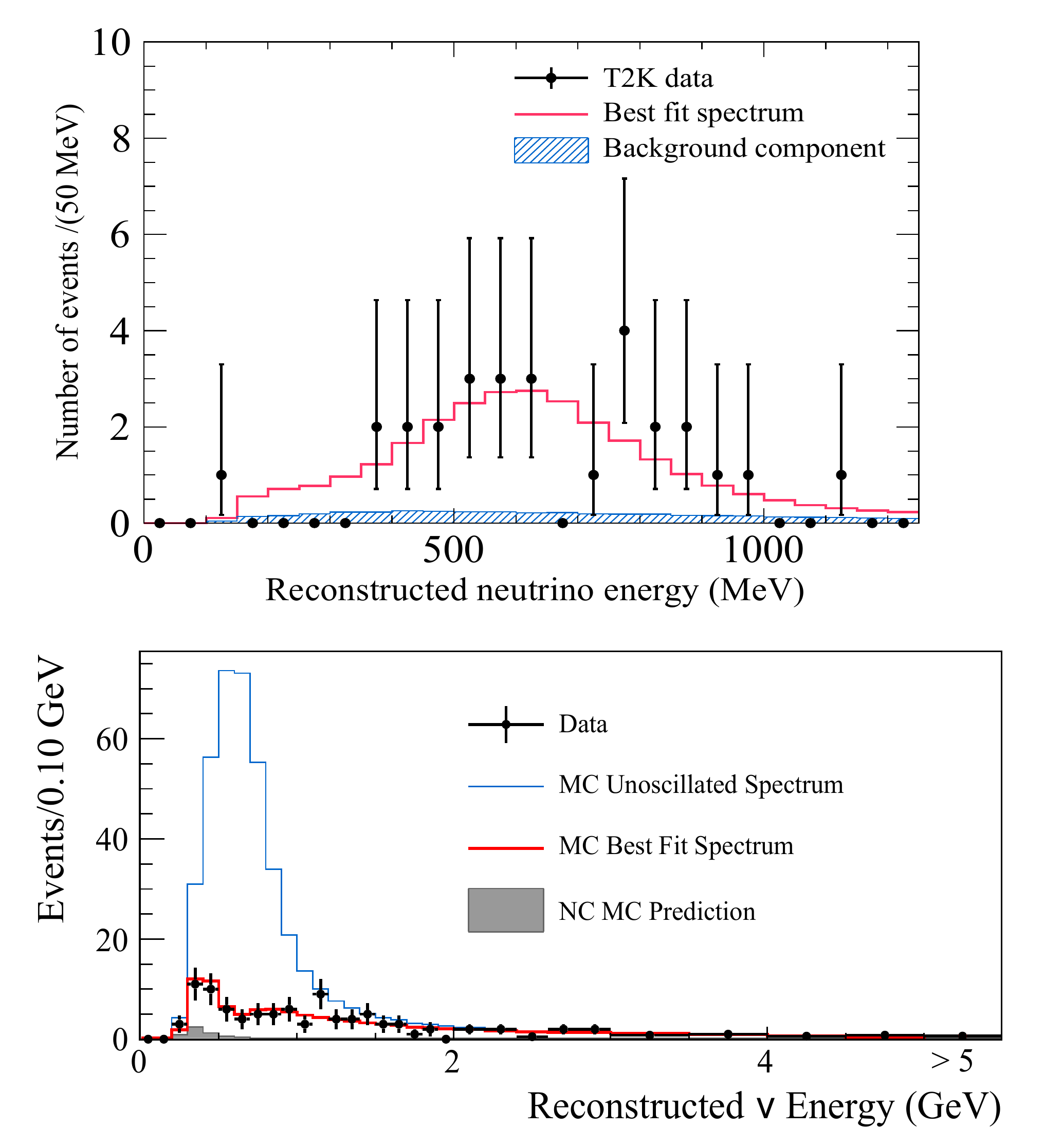}
\par\end{centering}
\caption{\label{fig:t2k_events} Top: Reconstructed energy spectrum of T2K $\nu_e$ candidate events \cite{Abe:2013hdq} with expected backgrounds (blue) and the best-fit prediction under $\nu_\mu \rightarrow \nu_e$ oscillation hypothesis (red). Bottom: Reconstructed energy spectrum of T2K $\nu_\mu$  candidate events \cite{Abe:2015t2k} with the expectation under no-oscillation hypothesis (blue), background (gray), and the best-fit prediction (red). Abbreviations: MC is Monte Carlo prediction. NC is the neutral current background contribution.
} 
\end{figure}

The T2K 
experiment in Japan, running since 2010,  has analyzed 
data from integrated exposure of $6.6\times 10^{20}$ protons
with the beam polarity in the neutrino mode. 
The top plot in Fig.~\ref{fig:t2k_events} shows the energy 
spectrum of  28  events identified as $\nu_e$  in T2K \cite{Abe:2013hdq}.
The background  expectation for these data from beam $\nu_e$ contamination
and neutral current were   3.2 and 1.0 events, respectively.  
 This observation  conclusively establishes the presence of
 $\nu_\mu \rightarrow \nu_e$ oscillations  driven by  $\Delta m^2_{31}$
 at $>7\sigma$ confidence level.
 The bottom panel of the 
figure shows the energy spectrum of the T2K  $\nu_\mu$ candidate 
events \cite{Abe:2015t2k} compared to no oscillation expectation. 
The  deficit of $\nu_\mu$ in T2K is 
consistent with maximal mixing angle $\theta_{23} = \pi/4$. A combined fit 
using both appearance and disappearance data results in  
$\sin^2\theta_{23} \approx 0.52\pm0.07$ with a small dependence on the assumed 
mass hierarchy.


T2K has an excess of $\nu_e$ events
compared to the prediction for $\delta_{CP} =0$ and 
the best fit value for $\sin^2 \theta_{13}$ from reactor data. 
 While T2K data alone cannot provide statistically significant 
 constraints on  $\delta_{CP}$, the addition of the 
reactor measurement  gives a preference for the negative 
values of $\delta_{CP}$ with the best-fit point near  
$-\pi/2$ (Fig. \ref{fig:deltacp_glob}). 
 Future addition of  antineutrino data from  T2K 
 as well as the NO$\nu$A data should further constrain
 the parameter range.  

The NO$\nu$A (NuMI Off-axis $\nu_e$ Appearance) experiment 
uses the Fermilab Main Injector 120 GeV proton beam. 
NO$\nu$A also follows the two-detector approach -- both off-axis at 14 mrad --
 with the near 
detector (at 1 km)  characterizing the neutrino beam 
close to the production 
point and the far (at 810 km)  measuring the
 flavor composition after oscillations. 
The  detectors use segmented liquid
 scintillator and are structurally and functionally
 identical with the near and far  active masses of 0.3 and 14 kton, 
respectively. 
The NO$\nu$A collaboration reported their first measurement of 
$\nu_e$ appearance 
and $\nu_\mu$ disappearance 
signals in the summer of 2015 \cite{Adamson:2016tbq}
after 400 kW beam operation for about a  year.   Depending on the 
event reconstruction method, the experiment has found  6 to 11 $\nu_e$ 
events with background expectation 
of $\sim 1$ event.  The $\nu_e$ event rate appears to be consistent
 with  $\delta_{CP} < 0$ with normal hierarchy and 
the disappearance result is consistent with
 the current picture of maximum mixing.



The size of the Earth matter effect depends on the energy of the 
oscillating neutrino (Eq.~\eqref{eq3}). 
The T2K and NO$\nu$A experiments 
have the oscillation maximum at $\sim0.6$ and $\sim1.6$ GeV, respectively.
The resulting large difference between matter effects, 
since $N^{res}_e \propto 1/E_\nu$, provides 
complementarity between the two experiments. 
The increase (or suppression) of $P_{\mu e}$ 
 due to the matter effect can be either 
masked or enhanced  by the CP asymmetry, creating an 
ambiguity if the asymmetry due to matter is smaller than  CP.  
The ambiguity can be reduced by 
a combination of 
measurements from the two experiments 
which will also provide information on  the neutrino mass hierarchy and 
$\delta_{CP}$ \cite{Abe:2014tzr}. 



\section{Next-generation Experiments}~\label{sec:future}  

The next generation experiments fall in two categories: 
detectors for  high statistics atmospheric neutrino studies,
and  well-optimized experiments  using man-made accelerator and 
reactor neutrinos. 
In both cases the  detectors need to be very large with excellent energy resolution and 
particle identification, and   well-shielded from cosmic ray muons. 


\subsection{Atmospheric Neutrinos} 

The  next-generation atmospheric neutrino 
experiments will focus on resolving the neutrino mass hierarchy (MH) 
~\cite{Franco:2013in,Ribordy:2013xea,Winter:2013ema,Ge:2013ffa,Capozzi:2015bxa}.
In the normal (inverted) hierarchy 
 there is a resonant effect for $\nu_\mu\to\nu_e$ ($\bar\nu_\mu\to\bar\nu_e$)
 at energy 
$E_{\nu}\sim5$ GeV and zenith angle $\cos\theta \sim -0.85$ 
(corresponding to traversal through the outer core of the Earth).
The effect can be measured through a large distortion in $P_{\mu\mu}$ 
and $P_{\bar\mu\bar\mu}$. 
Very large statistics (and correspondingly large detectors) 
are needed because of the limited solid angle and 
 the atmospheric flux power law spectrum, $\sim1/E^{2.7}$, in this energy range. 
The measurement requires excellent zenith angle resolution and 
an energy threshold  $\lesssim 5~{\rm GeV}$, relatively low for 
very large detectors optimized for very high
 energies ($>1~ {\rm TeV}$) such as IceCube. 
The capability
to differentiate positive and negative muons and muon tracks from showers
would allow enhanced sensitivity.  Muon charge can be determined either by 
magnetizing the detector or by utilizing the difference in absorption and 
decay  of $\mu^+$ and $\mu^-$ in dense materials. The latter requires very 
low energy threshold to be able to measure muon decay. 

 The proposed next-generation experiments include 
PINGU~\cite{Pingu}, 
ORCA 
~\cite{Katz:2014tta}, INO 
~\cite{Ahmed:2015jtv},
and Hyper-Kamiokande~\cite{Abe:2014oxa}.  
PINGU is a proposed multi-megaton high density array as an upgrade for IceCube
at  the South Pole Station. 
ORCA is proposed as a  deep-sea neutrino telescope 
in the Mediterranean as part of  KM3NeT  (multi-km$^3$ neutrino telescope).
INO is a planned to be an underground magnetized iron calorimeter with $\sim50$ kton mass and
1.5 Tesla magnetic field in southern India.
 Hyper-Kamiokande~\cite{Abe:2014oxa} is a 
proposed giant water Cherenkov detector 
 near the site of  the current Super-Kamiokande 
detector in Japan. 
 These experiments are expected to be  limited by systematics because of 
the energy and angular resolution needed. 
Their capability to distinguish between the two mass hierarchies, 
which is correlated to $\theta_{23}$,  
 can reach above 3$\sigma$  independent of the CP phase given enough exposure.  As remarked in 
Sec.~\ref{sec:current}, there is a large ambiguity  between the CP phase and 
the mass hierarchy in current accelerator based experiments.  This 
can be further reduced by combining with  atmospheric neutrino 
data~\cite{Ghosh:2013zna}.   

The oscillations patterns for atmospheric neutrinos also exhibit  
dependence on $\delta_{CP}$, predominantly at low energies (below 
a few GeV), raising a possibility of obtaining a constraint of 
this parameter complimentary to CP-violation searches with accelerator 
beams~\cite{Abe:2011ts} if the mass hierarchy is well-known. 
A multi-Mton scale detector with low energy detection threshold ($<$0.5 GeV) 
 and good angular and energy resolution would, however, be required to 
realize a measurement of this phase with high degree of 
significance~\cite{Razzaque:2014vba} using atmospheric neutrinos.  







\subsection{Reactor Neutrino Oscillations}

\begin{figure}[H]
\begin{centering}
\includegraphics[width=0.95\textwidth]{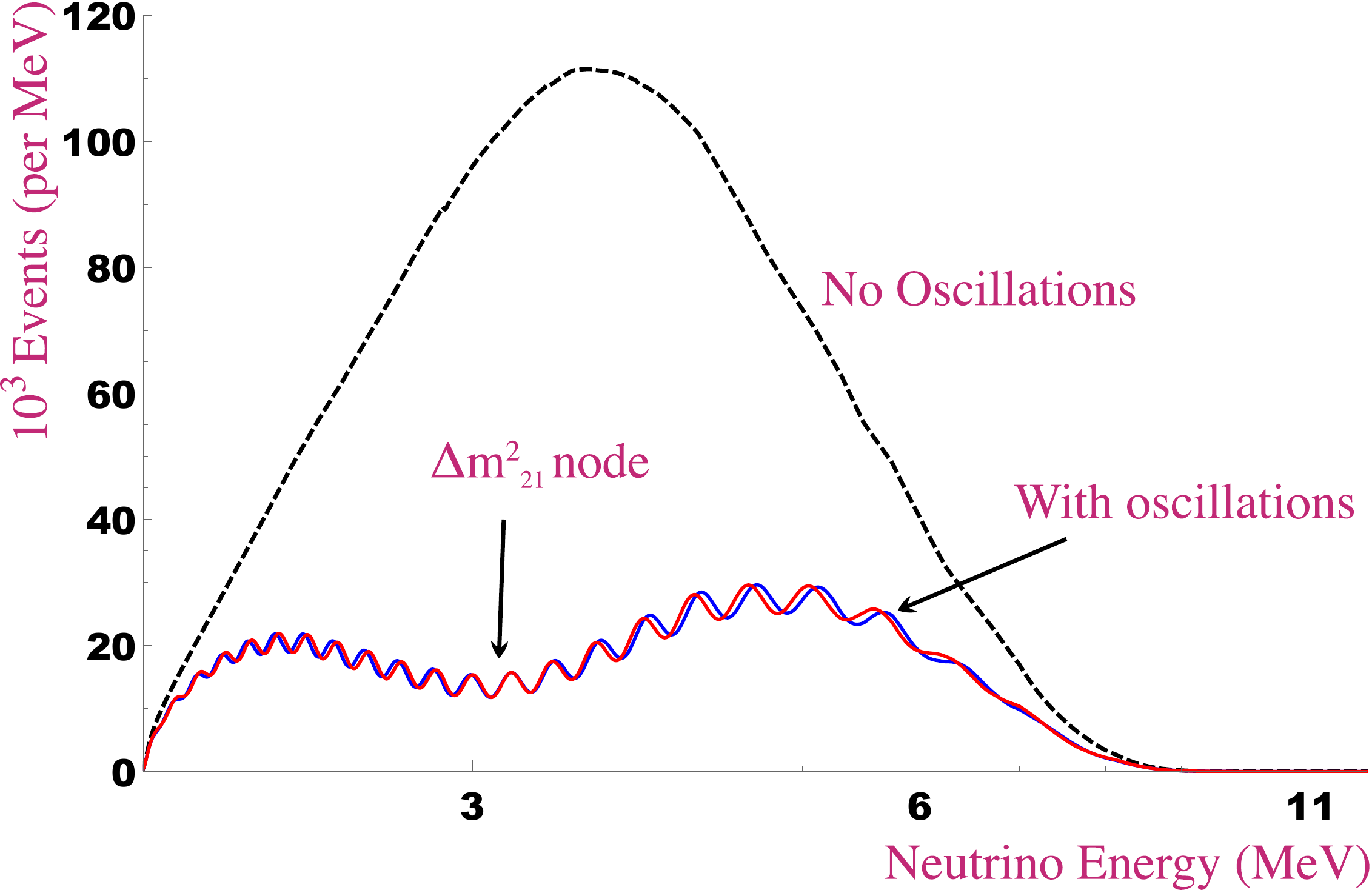}	
\par\end{centering}
\caption{\label{fig:reactor}  Reactor antineutrino spectrum is shown at 52.5 km
 for  no oscillations (dashed)\cite{Huber:2011wv,Mueller:2011nm}, 
and for $\Delta m^2_{32} = +2.4\times 10^{-3} {\rm eV^2}$ (red) for NH
and $-2.4\times 10^{-3} {\rm eV^2}$(blue) for IH.  
The event rate is normalized to 3000 kton$\cdot$GW$\cdot$year. 
To demonstrate the principle of the experiment no energy smearing is included. }
\end{figure}

The next generation reactor experiment will be optimized to resolve 
the mass hierarchy by detecting the $\sim \pm 3\%$ shift (Sec.~\ref{sec:pheno})
 in energy dependent oscillations in the $2$--$8 {\rm ~MeV}$  range.   
As usual  the choice of the  baseline distance is important. 
The fast oscillations due to the last two terms in Eq.~\eqref{eq:medreact}
 ($\Delta_{31}$ and $\Delta_{32}$) will exhibit the shift at any  distance, however at 
short distances (defined by $\Delta_{21}\ll \pi/2 $), 
the shift is  ambiguous with the measurement of 
$\Delta m^2_{31}$ itself, and at long 
distances ($\Delta_{21} \gg \pi/2$) 
the oscillations are too fast to be observable with achievable energy resolution.  
Using Eq. \ref{eq:medreact} we can write the 
asymmetry $P^{IH}(\bar\nu_e\to\bar\nu_e) - P^{NH}(\bar\nu_e\to\bar\nu_e)$ 
in obvious notation: 
\begin{equation} 
P^{IH}(\bar\nu_e\to\bar\nu_e)-P^{NH}(\bar\nu_e\to\bar\nu_e)= 
\sin^22\theta_{13} \cos^2\theta_{12} \sin 2 \Delta_{32} 
\sin 2 \Delta_{21}. 
\end{equation} 
At $\Delta_{21} = \pi/2$ or the maximum of solar oscillations, the asymmetry 
between IH and NH vanishes.  For $\Delta_{21} <\pi/2$, 
 the nodes of $P_{\bar e \bar e}$ 
oscillations are shifted to lower (higher) energies for normal (inverted) hierarchy, but 
for $\Delta_{21}>\pi/2$, the converse is true.  
This phenomenology allows us to choose $L$ so  that the 
$\Delta_{21} = \pi/2$ node is around $\sim 3$ MeV or the maximum of the reactor spectrum 
to avoid  ambiguities~\cite{Qian:2012xh}.
This node can be shown to be  at $L = {2\pi E/\Delta m^2_{21}} \approx 50{\rm~km}$. 
The oscillations on either side of the node at $\sim 3 {\rm~MeV}$  
can be compared in a single detector
to resolve for IH versus NH, reducing dependence on the absolute energy calibration
 (Fig. \ref{fig:reactor}). 
 Nevertheless, the detector must have sufficient statistics ($\sim 10^5$ events 
after oscillations), 
energy resolution($\lesssim 3 \%$ at 1 MeV),
and linearity ($\lesssim 1\%$)  across the reactor energy range. 
Otherwise,   low-energy
oscillation pattern  will  be smeared out. 
The detector also must be equidistant from all reactor cores 
so that the oscillation pattern is not obscured.  
The  large event sample from such an arrangement would also enable precision ($\sim 1\%$) 
measurements of $\sin^22\theta_{12}$, $\Delta m^2_{21}$, and $\Delta m^2_{32}$.


There are two proposed next generation reactor
experiments: the Jiangmen Underground Neutrino Observatory (JUNO) 
\cite{Djurcic:2015vqa,An:2015jdp} in China 
and  RENO-50 \cite{Kim:2014rfa} 
in South Korea. JUNO (RENO-50) design calls for a 20 kton (18 kton) 
fiducial mass liquid scintillator 
detector placed 700 m (900 m) underground, 52.5 km (47 km)  away from a 
set of reactors with total thermal 
power of $\sim 36 {\rm~GW}$ ($\sim 16 {\rm~GW}$). 
At about 50 km, the  IBD    
rate is $\sim0.1 (0.3) 
{\rm~day^{-1}kton^{-1}GW^{-1}}$ with (without) oscillations 
indicating  a need for 
$\sim$3000 ~kton$\cdot$GW$\cdot$year of exposure.

The needed energy 
resolution and linearity   lead to requirements on  the scintillator brightness, 
attenuation length, photo-detector coverage, and quantum efficiency so that the 
photo-electron  yield is $\gtrsim 1000$  per MeV.  
With the required energy resolution JUNO will achieve $>3\sigma$ separation of 
the two mass hierarchies with $\sim$6 years of operation\cite{An:2015jdp}.

\subsection{Accelerator Neutrino Oscillations}


Precision studies of neutrino oscillations with accelerators will 
focus on the search for CP violation with an  
optimized combination of neutrino beam energy, distance to the detector, 
and  detector technology.  
The preferred 
mode for future accelerator studies is $\nu_\mu \to \nu_e$ 
and the charge conjugate $\bar\nu_\mu \to \bar\nu_e$ (See Sec.\ref{sec:pheno});  
  this requires 
production of an intense $\nu_\mu ~(\bar\nu_\mu) $  beam. 
Using the known parameter values the   
 expected size of the signal at first maximum is
$\sin^2\theta_{23}\sin^2 2 \theta_{13} \approx 5\%$,  well above the 
the contamination  $\nu_e$ background of $\sim~1\%$ in a horn focused beam. 
The large expected signal makes background reduction as described in 
Sec.~\ref{sec:current}  a secondary concern. And therefore a broadband 
beam can be utilized to maximize statistics and the dynamic range. 
Furthermore alternative extremely low background methods using very 
pure intense beams of $\nu_e$ and $\bar\nu_e$  
using $\mu$ decay such as the
Neutrino Factory~\cite{Choubey:2011zzq, Geer:1997iz}  or decays of radioactive beta 
beams~\cite{Wildner:2014kia} are no longer favored.  

The asymmetry $A^{\mu e}$ has 
contributions from CP violation of $\le~32\%$, 
and from the matter effect which grows
as a function of the oscillation energy (or the length of baseline) as 
$\sim~\pm~2 N_e/N^{res}_e$, with the sign dependent on MH.  
At the first oscillation maximum, $\Delta_{31}\cong \Delta_{32} = \pi/2$, 
the asymmetry due to matter effect exceeds the maximum from 
CP violation at $E_\nu \sim~{\rm 2.3~GeV}$ or  
$\sim 1200{\rm~km}$~\cite{Bass:2013vcg,::2013kaa}.  
 Therefore, two different strategies for experimental optimization exist:  
distances of $\lesssim500{\rm~km}$ would   allow measurement of CP violation with low 
dependence on MH, while distances of $\gtrsim 1200 {\rm~km}$ 
would allow resolution of MH and also be sensitive to CP violation.  
A minimum distance is determined from the requirement that 
$E_\nu \gtrsim 0.5 {\rm ~GeV}$, a condition that ensures high enough 
probability for $\nu$ and $\bar{\nu}$ interactions 
and good particle reconstruction and identification. 
An additional consideration is to  obtain events at the second 
oscillation maximum,
$\Delta_{32}=3\pi/2$ where CP effects are a factor of 3  larger, but event rate 
is one-ninth as large due to kinematics. 
A broadband beam designed for baselines of $\gtrsim 1200 {\rm km}$  
will  also
have sensitivity at second maximum~\cite{Diwan:2003bp,::2013kaa}, an important 
consideration in resolving ambiguities.  

The expected event rate for a broadband beam
 can be calculated to lowest order by integrating 
the muon neutrino flux, $\Phi(E_\nu)\cong C/L^2$ with L in km and  
$C \sim 10^{17}~{\rm~\nu_\mu/m^2/GeV/MW/yr}$ 
with the appearance probability  
$P_{\mu e}$, and the charged current cross section 
$\sigma(E_\nu)\cong 0.7\times 10^{-38} E_\nu {\rm~cm^2/GeV/nucleon}$.  
 For vacuum oscillations 
 the event rate is found to be independent of the 
baseline distance because of the increase of oscillation probability and the
 cross section so that 
$N_{\nu_\mu\to \nu_e}(L)\sim \mathcal{O}(20) {\rm events/(kt\cdot MW\cdot yr)}$, with  the 
rate for $\bar\nu$ approximately $1/3$ of $\nu$. 
For off-axis beams and distances $\lesssim 500{\rm~km}$, 
the event yield and the ratio of $\bar\nu/\nu$ events 
is  smaller.  
For most modern high energy 
accelerators, the available beam power is limited to 
$\lesssim 1 {\rm ~MW}$\cite{Henderson:2008zz}, therefore 
 detectors with efficient mass (mass times efficiency)
 of $\gtrsim 50 {\rm~kton}$ are needed 
independent of distance to obtain a few hundred $\nu_e$ 
appearance events.  
At such large scales, water Cherenkov and liquid argon
 time projection chambers 
are considered  cost
effective  technologies~\cite{Barger:2007yw,Rubbia:2010zz}. 
At low energies ($\lesssim 1 {\rm~GeV}$)-- 
 distances of $\lesssim 500{\rm~km}$ --
the charged current $\nu_e$ cross section is dominated by 
quasi-elastic interactions with low multiplicity  final states,   
well-reconstructed by water Cherenkov detectors. 
However, at higher energies 
($\gtrsim 2{\rm~GeV}$) --  distances of $\gtrsim 1000{\rm~km}$ -- 
charged current 
events with multiple final state particles must be 
reconstructed to retain high efficiency;
a high granularity detector such as a 
liquid argon time projection chamber 
 is therefore preferred.



The Hyper-Kamiokande  (Hyper-K) experiment~\cite{Abe:2011ts,Abe:2014oxa} in  Japan
with a baseline of 295 km  and 
the Deep Underground Neutrino Experiment (DUNE) at the Long-Baseline 
Neutrino Facility (LBNF) in the U.S. with a baseline of 
$\sim1300{\rm ~km}$~\cite{Adams:2013qkq,Rubbia:2010zz,Acciarri:2015uup}
have chosen the two complementary approaches as outlined above.  

\begin{figure}[H]
\begin{centering}
\mbox{\includegraphics[width=0.48\textwidth]{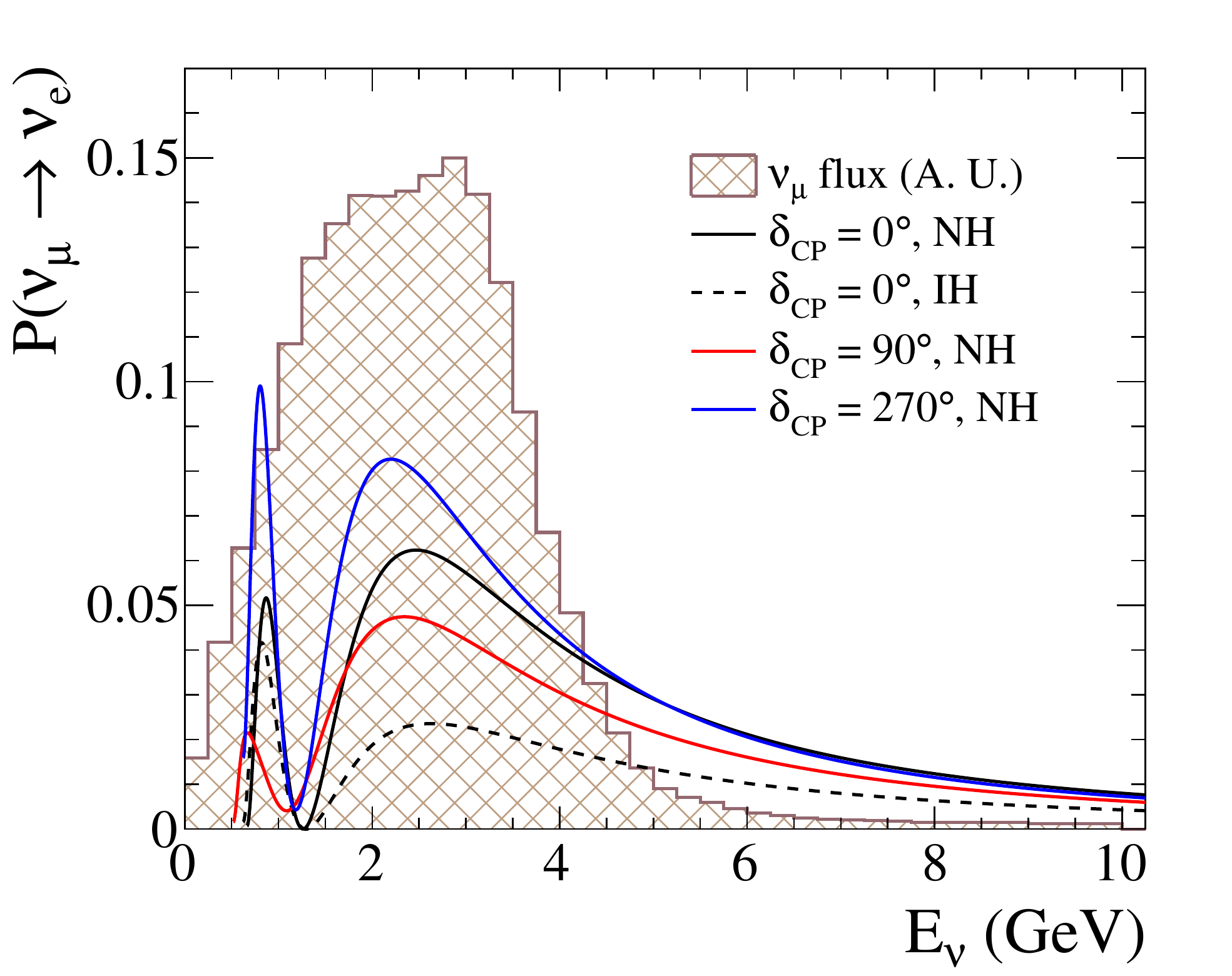}
\includegraphics[width=0.48\textwidth]{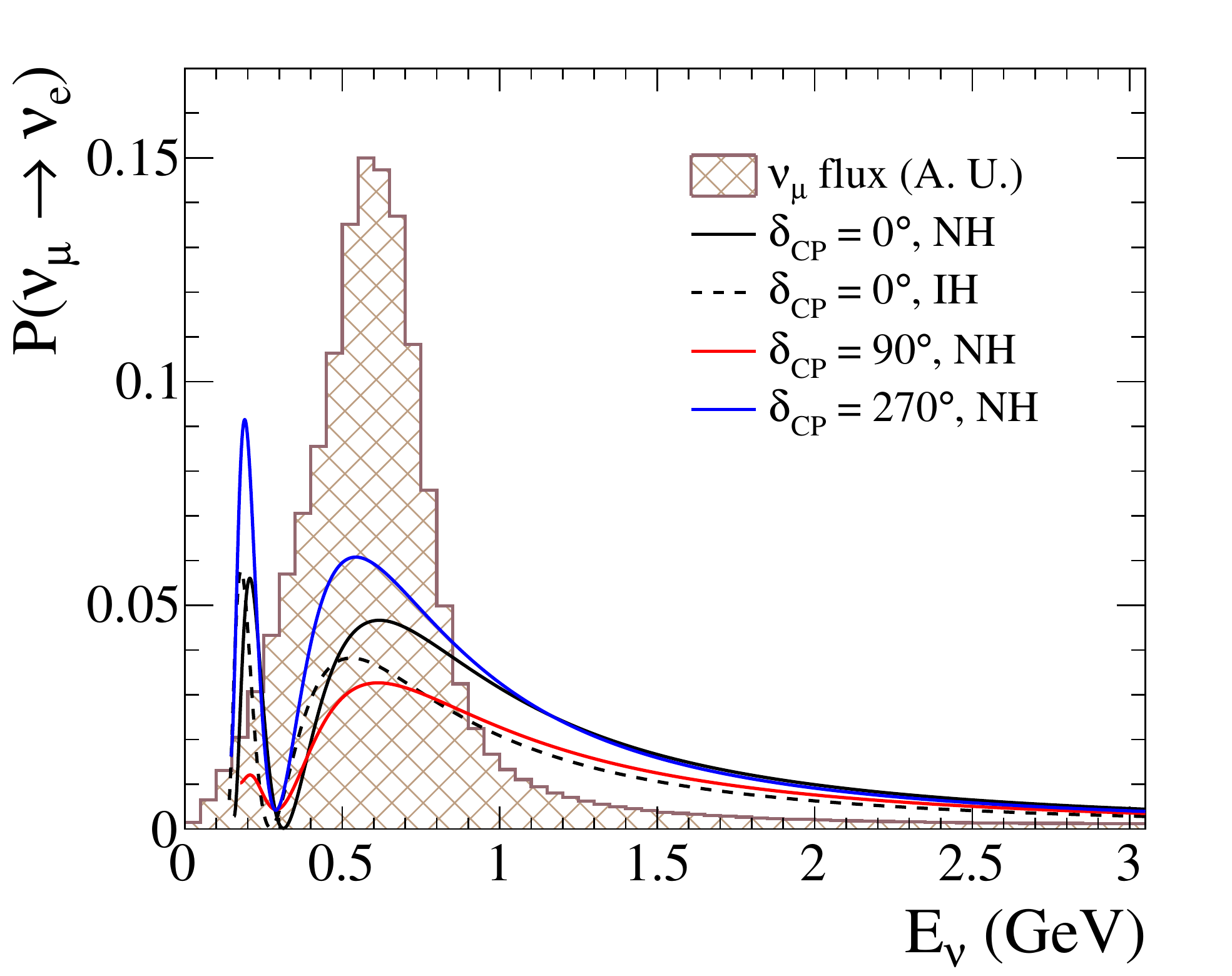} }
\end{centering}
\caption{\label{fig:flux}
$P(\nu_\mu \to \nu_e)$ plotted as a function of energy for 1300 km and 295 km
with the neutrino spectrum designed for DUNE (left) and Hyper-K (right) 
 in arbitrary units.  The probability for antineutrinos will be modified 
approximately by $\delta_{CP} \to -\delta_{CP}$ and NH$\to$IH. The 
antineutrino spectrum shape  is approximately the same. } 
\end{figure}

Hyper-K is planned to be a water Cherenkov detector with 
560 kton (1 Mton) fiducial (total) mass  near
 the current Super-K site in western Japan at a 
depth of $\sim 650{\rm~m}$. 
The Hyper-K detector will be placed in the same 
$44$ mrad off-axis neutrino beam matched to the first oscillation
node  (Fig.~\ref{fig:flux}) as T2K
 but with a much larger detector and 
beam  power of $\sim 1.3{\rm~ MW}$. 
Hyper-K will also  have a complex of near detectors to 
 monitor and measure the beam to predict background and signal 
rates at the far detector. 


DUNE consists of a horn-produced broad band beam with 60-120 GeV protons with
 beam-power of 
$\sim$1.2~MW from Fermilab, 40 kton fiducial volume liquid argon time 
projection far detector  $\sim$1450 m underground at 
Sanford Underground Research Laboratory
in  South Dakota,  and high-resolution near 
detector~\cite{Acciarri:2015uup}. 
The baseline of DUNE~\cite{Adams:2013qkq, Bass:2013vcg} is $\sim$1300~km which is optimized to
 measure the matter effect and CP violation simultaneously.  
The DUNE broad-band flux (Fig.~\ref{fig:flux}) is designed to 
cover the first ($\sim 2.5 {\rm ~GeV}$) and second ($\sim 0.8{\rm ~GeV}$) 
oscillation nodes sufficiently so that the CP phase can be measured
using the distortion of the energy spectrum as well as the $\nu, \bar\nu$ 
asymmetry.



The liquid argon time projection 
technology~\cite{Willis:1974gi,Nygren:1976fe,Chen:1976pp,lartpc}, 
chosen for DUNE,  could be implemented
by two approaches: 
the single phase in which the drifting charge is detected by   electrodes
 in the liquid argon~\cite{Adams:2013qkq}
or the double phase in which the charge is drifted to the surface
 of the liquid argon and 
amplified in the gas phase before detection~\cite{::2013kaa}. 

\begin{figure}[H]
\begin{centering}
\includegraphics[width=0.95\textwidth]{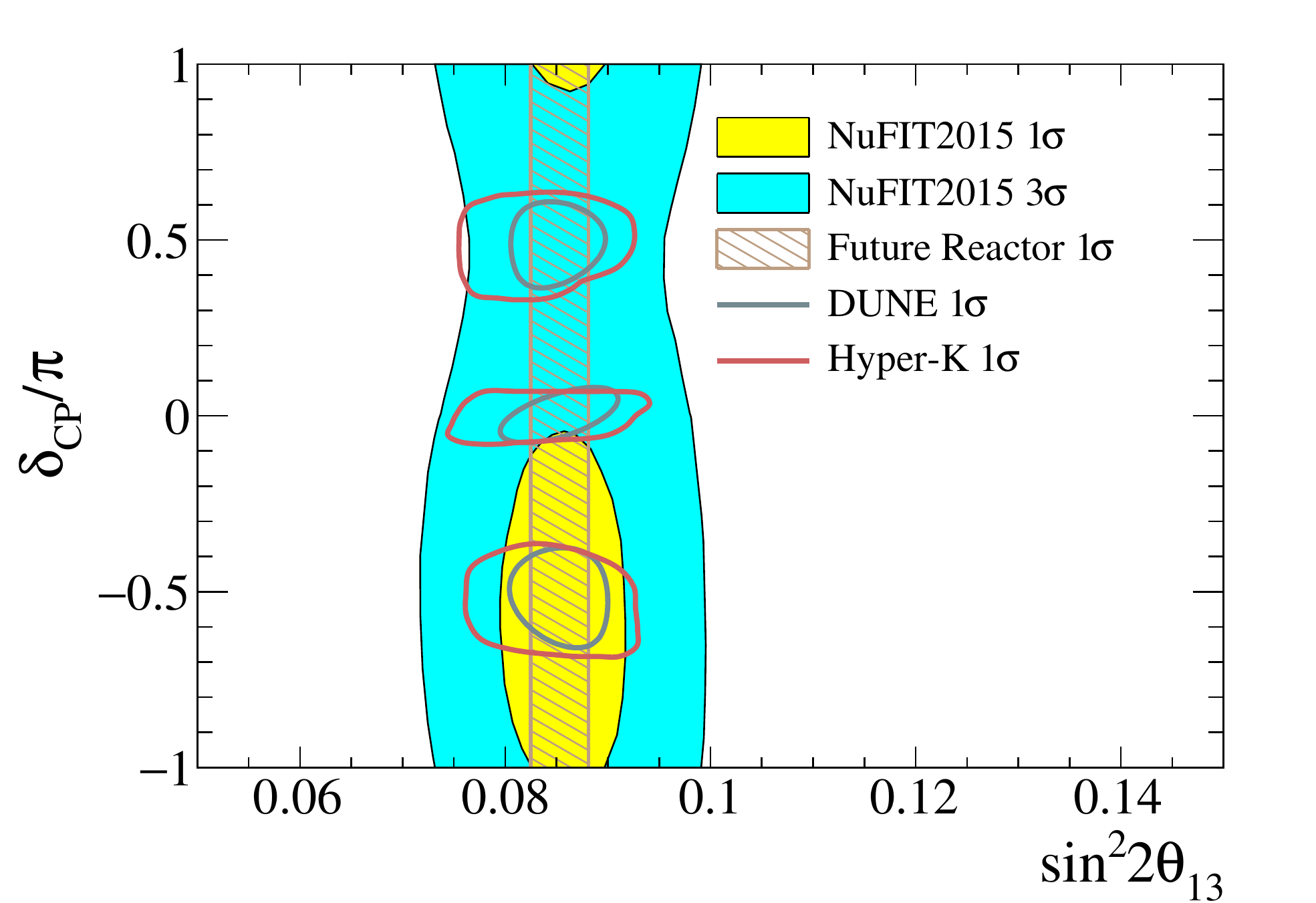}
\par\end{centering}
\caption{\label{fig:deltacp_glob} 
Current measurements and future goals in the $\delta_{CP}$ and $\sin^2 2 \theta_{13}$ plane. 
 The current best fit in yellow (cyan) for $1\sigma$ ($3\sigma$) C.L. is from a combination 
of T2K and reactor measurements~\cite{Gonzalez-Garcia:2014bfa}. Brown vertical band shows the future reactor measurement
 dominated by Daya Bay.  The blue-gray (maroon) contours show measurements 
that could result from DUNE (Hyper-K) at $\delta_{CP}$ values of $\pm 90^\circ$ and $0^\circ$. 
The Hyper-K and DUNE projections are taken from Ref.~\cite{Abe:2015zbg} 
and Ref.~\cite{Acciarri:2015uup}, respectively. Hyper-K and DUNE do not 
have identical treatment of  the correlation to $\theta_{23}$. 
 For Hyper-K the mass hierarchy is 
assumed to be known; for DUNE the measurement can be done with the same data.}
\end{figure}

The Hyper-K and DUNE experiments require precise prediction of 
   $\nu_e$  signal and background spectra in the far detector.  
 The prediction  is calculated  by extrapolating the measured 
event spectrum from the near detector, using the known beam geometry 
and constraints on  the cross sections, and   near and far detector efficiencies.  
The systematic errors associated with neutrino nucleus cross sections 
require  further measurements and modeling ~\cite{Mosel:2016ar}. 
Since the maximum expected asymmetry from CP violation is $32\%$, the requirement 
on the allowed systematic error is less than a few percent 
so that a $5\sigma$ effect 
can be measured with $\mathcal{O}(1000)$ events.  A joint fit to  
the four spectra -- $\nu_\mu$, $\bar\nu_\mu$, $\nu_e$, $\bar\nu_e$ --
 is expected to cancel some 
uncertainties, and the remaining largest contribution  is expected to come 
from the relative energy scale for $\nu_e$ and
 $\nu_\mu$  events~\cite{Proceedings:2015zqa}. 
For DUNE, if the matter and CP asymmetries have the same sign, the combined 
effect is very large
and the mass hierarchy will be determined to $>5 \sigma$
 within a few months of running;
 in case of the opposite,  exposure of a few years
 is needed~\cite{Adams:2013qkq}.  
Figure~\ref{fig:deltacp_glob} shows the current global fit 
for $\delta_{CP}$ versus $\sin^2 2 \theta_{13}$ while marginalizing over all other 
oscillation parameters~\cite{Gonzalez-Garcia:2014bfa,Forero:2014bxa} including the 
reactor constraint on $\theta_{13}$. The figure also
shows the expectation at $\delta_{CP} = 0$~and~$\pm \pi/2$
 for Hyper-K (DUNE) assuming 10 yrs of exposure at 1.3 MW
(1.2 MW)~\cite{Abe:2015zbg,Acciarri:2015uup}. 
Note that the Hyper-K and DUNE expectations do not include the $\theta_{13}$ constraint.
The independent measurements of $\theta_{13}$ from accelerators and reactors 
can be compared to test for new physics~\cite{Qian:2013ora}.



\section{Conclusion }\label{sec:summary}

We have reviewed the current status of the physics of long-baseline 
neutrino oscillations 
with emphasis on the experimental technique.  
The picture of 3-flavor oscillations has now been established 
by observation of  neutrino oscillations with solar, atmospheric, 
reactor, and accelerator neutrinos.  
Remarkably, a consistent picture of  mass differences 
 and mixing has emerged 
so that extremely precise experiments,  using well-characterized 
and controlled terrestrial sources of neutrinos using reactors and accelerators, are now feasible. 
The main objectives of such an experimental  program are the determination of potentially large 
CP violation in the neutrino sector, resolution of the mass ordering, and 
measurement of the mixing parameters with greater precision.  
The precision and redundancy of measurements will allow constraints on new physics beyond 
the 3-flavor model as well as neutrino mass and mixing models.

The writing of this review was supported by the US Department of Energy for 
M.V.D and X.Q., the Swiss National Science Foundation (SNF) for A.R., and
the French National Centre for Scientific Research (CNRS) for V.G..


\bibliographystyle{arnuke_revised}
\bibliography{LBN_review}{}

\begin{thebibliography}{100}

\bibitem{brown1978}
Brown LM,
\newblock Physics Today 31:23 (1978).

\bibitem{fermi1934}
Fermi E,
\newblock Il Nuovo Cimento 11:1 (1934).

\bibitem{Reines1956}
Reines F, Cowan CL,
\newblock Nature 178:446 (1956).

\bibitem{Cowan1956}
Cowan CL, Reines F, Harrison FB, Kruse HW, McGuire AD,
\newblock Science 124:103 (1956).

\bibitem{winter2000}
Winter K,
\newblock {\em Neutrino Physics} (Cambridge Uni. Pr., 2000).

\bibitem{kolb-turner}
Kolb EW, Turner MS,
\newblock {\em The Early Universe} (Westview Pr., 1990).

\bibitem{Agashe:2014kda}
Part. Data Group, Olive KA, et~al.,
\newblock Chin. Phys. C38:090001 (2014).

\bibitem{theory-white-paper}
Mohapatra RN, et~al.,
\newblock Rept. Prog. Phys. 70:1757 (2007), hep-ph/0510213.

\bibitem{ponte57-1}
Pontecorvo B,
\newblock Zh. Eksp. Theor. Phys. 33:549 (1957).

\bibitem{ponte1}
Pontecorvo B,
\newblock Sov. Phys. JETP 6:429 (1958).

\bibitem{ponte67-2}
Pontecorvo B,
\newblock Zh. Eksp. Theor. Phys. 53:1717 (1967).

\bibitem{ponte2}
Pontecorvo B,
\newblock Sov. Phys. JETP 26:989 (1967).

\bibitem{Maki}
Maki Z, Nakagawa M, Sakata S,
\newblock Prog. Theor. Phys. 28:247 (1962).

\bibitem{twonus-1}
Danby G, Gaillard JM, Goulianos KA, Lederman LM, Mistry NB, et~al.,
\newblock Phys. Rev. Lett. 9:36 (1962).

\bibitem{smirnov1}
Akhmedov EK, Smirnov A{\relax Yu},
\newblock Phys. Atom. Nucl. 72:1363 (2009), 0905.1903.

\bibitem{Commins:1983ns}
Commins ED, Bucksbaum PH,
\newblock {\em {Weak Interactions of Leptons and Quarks}} (Cambridge Univ. Pr.,
  1983).

\bibitem{Wolfenstein:1977ue}
Wolfenstein L,
\newblock Phys. Rev. D17:2369 (1978).

\bibitem{Mikheev:1986gs}
Mikheev SP, Smirnov A{\relax Yu},
\newblock Sov. J. Nucl. Phys. 42:913 (1985),
\newblock [Yad. Fiz.42,1441(1985)].

\bibitem{Barger:1980tf}
Barger VD, Whisnant K, Pakvasa S, Phillips RJN,
\newblock Phys. Rev. D22:2718 (1980).

\bibitem{Marciano:2008zz}
Marciano WJ, Mori T, Roney JM,
\newblock Ann. Rev. Nucl. Part. Sci. 58:315 (2008).

\bibitem{Mihara:2013zna}
Mihara S, Miller JP, Paradisi P, Piredda G,
\newblock Ann. Rev. Nucl. Part. Sci. 63:531 (2013).

\bibitem{Otten:2008zz}
Otten EW, Weinheimer C,
\newblock Rept. Prog. Phys. 71:086201 (2008), 0909.2104.

\bibitem{Robertson:2012ib}
Haxton WC, Hamish~Robertson RG, Serenelli AM,
\newblock Ann. Rev. Astron. Astrophys. 51:21 (2013), 1208.5723.

\bibitem{Mueller:2011nm}
Mueller TA, et~al.,
\newblock Phys. Rev. C83:054615 (2011), 1101.2663.

\bibitem{Huber:2011wv}
Huber P,
\newblock Phys. Rev. C84:024617 (2011), 1106.0687,
\newblock [Erratum: Phys. Rev.C85,029901(2012)].

\bibitem{Gaisser:2002jj}
Gaisser TK, Honda M,
\newblock Ann. Rev. Nucl. Part. Sci. 52:153 (2002), hep-ph/0203272.

\bibitem{Honda:2011nf}
Honda M, Kajita T, Kasahara K, Midorikawa S,
\newblock Phys. Rev. D83:123001 (2011), 1102.2688.

\bibitem{Suzuki:2005id}
Suzuki Y, Nakahata M, Moriyama S, Koshio Y, eds.,
\newblock {\em Neutrino oscillations and their origin. Proceedings, 5th
  International Workshop, NOON2004, Tokyo, Japan, February 11-15, 2004}, 2005.

\bibitem{horns}
Dusseux JC, Pattison JBM, Ziebarth G,
\newblock The cern magnetic horn (1971) and its remote-handling system.,
\newblock CERN-72-11 June 1972.

\bibitem{Gonzalez-Garcia:2014bfa}
Gonzalez-Garcia MC, Maltoni M, Schwetz T,
\newblock JHEP 11:052 (2014), 1409.5439.

\bibitem{Forero:2014bxa}
Forero DV, Tortola M, Valle JWF,
\newblock Phys. Rev. D90:093006 (2014), 1405.7540.

\bibitem{Mena:2003ug}
Mena O, Parke SJ,
\newblock Phys. Rev. D69:117301 (2004), hep-ph/0312131.

\bibitem{Fukuda:2002uc}
Super-Kamiokande Collab., Fukuda Y, et~al.,
\newblock Nucl. Instrum. Meth. A501:418 (2003).

\bibitem{Fukuda:1998mi}
Super-Kamiokande Collab., Fukuda Y, et~al.,
\newblock Phys. Rev. Lett. 81:1562 (1998), hep-ex/9807003.

\bibitem{Hatakeyama:1998ea}
Kamiokande Collab., Hatakeyama S, et~al.,
\newblock Phys. Rev. Lett. 81:2016 (1998), hep-ex/9806038.

\bibitem{Clark:1997cb}
Clark R, et~al.,
\newblock Phys. Rev. Lett. 79:345 (1997).

\bibitem{kajita-annual-2014}
Kajita T,
\newblock Ann. Rev. Nucl. Part. Sci. 64:343 (2014).

\bibitem{Ashie:2004mr}
Super-Kamiokande Collab., Ashie Y, et~al.,
\newblock Phys. Rev. Lett. 93:101801 (2004), hep-ex/0404034.

\bibitem{K2K}
K2K Collab., Ahn MH, et~al.,
\newblock Phys. Rev. D74:072003 (2006).

\bibitem{MINOS}
MINOS Collab., Adamson P, et~al.,
\newblock Phys. Rev. Lett. 110:251801 (2013).

\bibitem{Abe:2013fuq}
T2K Collab., Abe K, et~al.,
\newblock Phys. Rev. Lett. 111:211803 (2013), 1308.0465.

\bibitem{Adamson:2014vgd}
MINOS Collab., Adamson P, et~al.,
\newblock Phys. Rev. Lett. 112:191801 (2014), 1403.0867.

\bibitem{Abe:2014ugx}
T2K Collab., Abe K, et~al.,
\newblock Phys. Rev. Lett. 112:181801 (2014), 1403.1532.

\bibitem{Adamson:2007gu}
MINOS Collab., Adamson P, et~al.,
\newblock Phys. Rev. D77:072002 (2008), 0711.0769.

\bibitem{Beavis-BNL-52459}
E889 Collab., Beavis D, et~al.,
\newblock Brookhaven National Laboratory Report No. BNL-52459, 1995
  (unpublished),
\newblock www.osti.gov No. 52878.

\bibitem{Abe:2011ks}
T2K Collab., Abe K, et~al.,
\newblock Nucl. Instrum. Meth. A659:106 (2011), 1106.1238.

\bibitem{PhysRevLett.110.181802}
Super-Kamiokande Collab., Abe K, et~al.,
\newblock Phys. Rev. Lett. 110:181802 (2013).

\bibitem{Yoshida:2013pva}
Yoshida J, et~al.,
\newblock JINST 8:P02009 (2013).

\bibitem{Agafonova:2015jxn}
OPERA Collab., Agafonova N, et~al.,
\newblock Phys. Rev. Lett. 115:121802 (2015), 1507.01417.

\bibitem{PhysRevD.89.051102}
OPERA Collab., Agafonova N, et~al.,
\newblock Phys. Rev. D 89:051102 (2014).

\bibitem{Adamson:2013whj}
MINOS Collab., Adamson P, et~al.,
\newblock Phys. Rev. Lett. 110:251801 (2013), 1304.6335.

\bibitem{Cleveland:1994er}
Cleveland BT, Daily T, Davis Jr. R, Distel J, Lande K, et~al.,
\newblock Nucl. Phys. Proc. Suppl. 38:47 (1995).

\bibitem{Cleveland:1998nv}
Cleveland BT, Daily T, Davis Jr. R, Distel JR, Lande K, et~al.,
\newblock Astrophys. J. 496:505 (1998).

\bibitem{Bahcall:2000nu}
Bahcall JN, Pinsonneault MH, Basu S,
\newblock Astrophys. J. 555:990 (2001), astro-ph/0010346.

\bibitem{Abdurashitov:1999zd}
SAGE Collab., Abdurashitov JN, et~al.,
\newblock Phys. Rev. C60:055801 (1999), astro-ph/9907113.

\bibitem{Hampel:1998xg}
GALLEX Collab., Hampel W, et~al.,
\newblock Phys. Lett. B447:127 (1999).

\bibitem{Fukuda:1996sz}
Kamiokande Collab., Fukuda Y, et~al.,
\newblock Phys. Rev. Lett. 77:1683 (1996).

\bibitem{Fukuda:1998fd}
Super-Kamiokande Collab., Fukuda Y, et~al.,
\newblock Phys. Rev. Lett. 81:1158 (1998), hep-ex/9805021,
\newblock [Erratum: Phys. Rev. Lett.81,4279(1998)].

\bibitem{Fukuda:2002pe}
Super-Kamiokande Collab., Fukuda S, et~al.,
\newblock Phys. Lett. B539:179 (2002), hep-ex/0205075.

\bibitem{Aharmim:2005gt}
SNO Collab, Aharmim B, et~al.,
\newblock Phys. Rev. C72:055502 (2005), nucl-ex/0502021.

\bibitem{Ahmad:2002jz}
SNO Collab., Ahmad QR, et~al.,
\newblock Phys. Rev. Lett. 89:011301 (2002), nucl-ex/0204008.

\bibitem{Bonventre:2013loa}
Bonventre R, LaTorre A, Klein JR, Orebi~Gann GD, Seibert S, Wasalski O,
\newblock Phys. Rev. D88:053010 (2013), 1305.5835.

\bibitem{MSW1}
Mikheyev SP, Smirnov AY,
\newblock Sov. J. Nucl. Phys. 42:913 (1986).

\bibitem{MSW2}
Mikheyev SP, Smirnov AY,
\newblock Sov. Phys. JETP 64:4 (1986).

\bibitem{MSW3}
Mikheyev SP, Smirnov AY,
\newblock Nuovo Cim 9C:17 (1986).

\bibitem{Renshaw:2013dzu}
Super-Kamiokande, Renshaw A, et~al.,
\newblock Phys. Rev. Lett. 112:091805 (2014), 1312.5176.

\bibitem{Eguchi:2002dm}
KamLAND Collab., Eguchi K, et~al.,
\newblock Phys. Rev. Lett. 90:021802 (2003), hep-ex/0212021.

\bibitem{Gando:2013nba}
KamLAND Collab., Gando A, et~al.,
\newblock Phys. Rev. D88:033001 (2013), 1303.4667.

\bibitem{Fogli:2011qn}
Fogli GL, Lisi E, Marrone A, Palazzo A, Rotunno AM,
\newblock Phys. Rev. D84:053007 (2011), 1106.6028.

\bibitem{Chooz1}
Chooz Collab., Apollonio M, et~al.,
\newblock Phys.Lett. B466:415 (1999), hep-ex/9907037.

\bibitem{Chooz2}
Chooz Collab., Apollonio M, et~al.,
\newblock Eur.Phys.J. C27:331 (2003), hep-ex/0301017.

\bibitem{PaloVerde}
Boehm F, Busenitz J, Cook B, Gratta G, Henrikson H, et~al.,
\newblock Phys.Rev. D64:112001 (2001), hep-ex/0107009.

\bibitem{dc}
Double-Chooz Collab., Abe Y, et~al.,
\newblock Phys.Rev.Lett. 108:131801 (2012), 1112.6353.

\bibitem{dayabay}
Daya-Bay Collab., An F, et~al.,
\newblock Phys.Rev.Lett. 108:171803 (2012), 1203.1669.

\bibitem{RENO}
RENO Collab., Ahn J, et~al.,
\newblock Phys.Rev.Lett. 108:191802 (2012), 1204.0626.

\bibitem{An:2013uza}
Daya-Bay Collab., An FP, et~al.,
\newblock Chin. Phys. C37:011001 (2013), 1210.6327.

\bibitem{An:2015rpe}
Daya-Bay Collab., An FP, et~al.,
\newblock Phys. Rev. Lett. 115:111802 (2015), 1505.03456.

\bibitem{minos1}
MINOS Collab., Adamson P, et~al.,
\newblock Phys.Rev.Lett. 107:181802 (2011), 1108.0015.

\bibitem{t2k}
T2K Collab., Abe K, et~al.,
\newblock Phys.Rev.Lett. 107:041801 (2011), 1106.2822.

\bibitem{Qian:2015waa}
Qian X, Vogel P,
\newblock Prog. Part. Nucl. Phys. 83:1 (2015), 1505.01891.

\bibitem{LSND}
Aguilar-Arevalo A, et~al.,
\newblock Phys. Rev. D64:112007 (2001).

\bibitem{miniboone}
MiniBooNE Collab., Aguilar-Arevalo A, et~al.,
\newblock Phys.Rev.Lett. 110:161801 (2013), 1207.4809.

\bibitem{anom}
Mention G, et~al.,
\newblock Phys. Rev. D83:073006 (2011).

\bibitem{Abazajian:2012ys}
Abazajian KN, et~al.,
\newblock (2012), 1204.5379.

\bibitem{unitarity_decay}
Antusch S, et~al.,
\newblock JHEP 0610:084 (2006).

\bibitem{Qian:2013ora}
Qian X, Zhang C, Diwan M, Vogel P,
\newblock (2013), 1308.5700.

\bibitem{Fukugita:1986hr}
Fukugita M, Yanagida T,
\newblock Phys. Lett. B174:45 (1986).

\bibitem{Abada:2006fw}
Abada A, Davidson S, Josse-Michaux FX, Losada M, Riotto A,
\newblock JCAP 0604:004 (2006), hep-ph/0601083.

\bibitem{Nardi:2006fx}
Nardi E, Nir Y, Roulet E, Racker J,
\newblock JHEP 01:164 (2006), hep-ph/0601084.

\bibitem{Pascoli:2006ie}
Pascoli S, Petcov ST, Riotto A,
\newblock Phys. Rev. D75:083511 (2007), hep-ph/0609125.

\bibitem{Pascoli:2006ci}
Pascoli S, Petcov ST, Riotto A,
\newblock Nucl. Phys. B774:1 (2007), hep-ph/0611338.

\bibitem{marciano:2001tz}
Marciano WJ,
\newblock (2001), hep-ph/0108181.

\bibitem{Parke:2005ev}
Parke SJ,
\newblock {Theta(13)},
\newblock in {\em Particles and nuclei : Seventeenth International Conference
  on Particles and Nuclei, Santa Fe, New Mexico, 23-30 October 2005}, 2005.

\bibitem{Bass:2013vcg}
Bass M, et~al.,
\newblock Phys. Rev. D91:052015 (2015), 1311.0212.

\bibitem{Diwan:2003bp}
Diwan MV, et~al.,
\newblock Phys. Rev. D68:012002 (2003), hep-ph/0303081.

\bibitem{Chizhov:1999he}
Chizhov MV, Petcov ST,
\newblock Phys. Rev. D63:073003 (2001), hep-ph/9903424.

\bibitem{Akhmedov:2006hb}
Akhmedov EK, Maltoni M, Smirnov A{\relax Yu},
\newblock JHEP 05:077 (2007), hep-ph/0612285.

\bibitem{Freund:2001pn}
Freund M,
\newblock Phys. Rev. D64:053003 (2001), hep-ph/0103300.

\bibitem{Diwan:2004bt}
Diwan MV,
\newblock Frascati Phys. Ser. 35:89 (2004), hep-ex/0407047,
\newblock [,89(2004)].

\bibitem{An:2013zwz}
Daya-Bay Collab., An FP, et~al.,
\newblock Phys. Rev. Lett. 112:061801 (2014), 1310.6732.

\bibitem{2013arXiv1312.4111S}
RENO Collab.,
\newblock (2015), 1511.05849.

\bibitem{Petcov:2001sy}
Petcov ST, Piai M,
\newblock Phys. Lett. B533:94 (2002), hep-ph/0112074.

\bibitem{Choubey:2003qx}
Choubey S, Petcov ST, Piai M,
\newblock Phys. Rev. D68:113006 (2003), hep-ph/0306017.

\bibitem{deGouvea:2005hk}
de~Gouvea A, Jenkins J, Kayser B,
\newblock Phys. Rev. D71:113009 (2005), hep-ph/0503079.

\bibitem{Learned:2006wy}
Learned J, Dye ST, Pakvasa S, Svoboda RC,
\newblock Phys. Rev. D78:071302 (2008), hep-ex/0612022.

\bibitem{Minakata:2007tn}
Minakata H, Nunokawa H, Parke SJ, Zukanovich~Funchal R,
\newblock Phys. Rev. D76:053004 (2007), hep-ph/0701151,
\newblock [Erratum: Phys. Rev.D76,079901(2007)].

\bibitem{Parke:2008cz}
Parke SJ, Minakata H, Nunokawa H, Funchal RZ,
\newblock Nucl. Phys. Proc. Suppl. 188:115 (2009), 0812.1879.

\bibitem{Zhan:2008id}
Zhan L, Wang Y, Cao J, Wen L,
\newblock Phys. Rev. D78:111103 (2008), 0807.3203.

\bibitem{Zhan:2009rs}
Zhan L, Wang Y, Cao J, Wen L,
\newblock Phys. Rev. D79:073007 (2009), 0901.2976.

\bibitem{Qian:2012xh}
Qian X, Dwyer DA, McKeown RD, Vogel P, Wang W, Zhang C,
\newblock Phys. Rev. D87:033005 (2013), 1208.1551.

\bibitem{Ciuffoli:2012iz}
Ciuffoli E, Evslin J, Zhang X,
\newblock JHEP 03:016 (2013), 1208.1991.

\bibitem{Ge:2012wj}
Ge SF, Hagiwara K, Okamura N, Takaesu Y,
\newblock JHEP 05:131 (2013), 1210.8141.

\bibitem{Li:2013zyd}
Li YF, Cao J, Wang Y, Zhan L,
\newblock Phys. Rev. D88:013008 (2013), 1303.6733.

\bibitem{Minakata:2004xt}
Minakata H, Smirnov A{\relax Yu},
\newblock Phys. Rev. D70:073009 (2004), hep-ph/0405088.

\bibitem{Altarelli:2010gt}
Altarelli G, Feruglio F,
\newblock Rev. Mod. Phys. 82:2701 (2010), 1002.0211.

\bibitem{Timmons:2015laq}
Timmons A,
\newblock (2015), 1511.06178.

\bibitem{Zhang:2015fya}
Daya-Bay Collab., Zhang C,
\newblock AIP Conf. Proc. 1666:080003 (2015), 1501.04991.

\bibitem{Aartsen:2014yll}
IceCube Collab., Aartsen M, et~al.,
\newblock Phys. Rev. D91:072004 (2015), 1410.7227.

\bibitem{Abe:2013hdq}
T2K Collab., Abe K, et~al.,
\newblock Phys. Rev. Lett. 112:061802 (2014), 1311.4750.

\bibitem{Abe:2015t2k}
T2K Collab., Abe K, et~al.,
\newblock Phys. Rev. D91:072010 (2015), 1502.01550.

\bibitem{Adamson:2016tbq}
NOvA, Adamson P, et~al.,
\newblock Phys. Rev. Lett. 116:151806 (2016), 1601.05022.

\bibitem{Abe:2014tzr}
T2K Collab., Abe K, et~al.,
\newblock PTEP 2015:043C01 (2015), 1409.7469.

\bibitem{Franco:2013in}
Franco D, Jollet C, Kouchner A, Kulikovskiy V, Meregaglia A, et~al.,
\newblock JHEP 04:008 (2013), 1301.4332.

\bibitem{Ribordy:2013xea}
Ribordy M, Smirnov AY,
\newblock Phys. Rev. D87:113007 (2013), 1303.0758.

\bibitem{Winter:2013ema}
Winter W,
\newblock Phys. Rev. D88:013013 (2013), 1305.5539.

\bibitem{Ge:2013ffa}
Ge SF, Hagiwara K,
\newblock JHEP 09:024 (2014), 1312.0457.

\bibitem{Capozzi:2015bxa}
Capozzi F, Lisi E, Marrone A,
\newblock Phys. Rev. D91:073011 (2015), 1503.01999.

\bibitem{Pingu}
IceCube-PINGU Collab., Aartsen M, et~al.,
\newblock (2014), 1401.2046.

\bibitem{Katz:2014tta}
KM3NeT Collab., Katz UF,
\newblock Submitted to: PoS  (2014), 1402.1022.

\bibitem{Ahmed:2015jtv}
ICAL Collab., Ahmed S, et~al.,
\newblock (2015), 1505.07380.

\bibitem{Abe:2014oxa}
Hyper-Kamiokande Working Group, Abe K, et~al.,
\newblock {A Long Baseline Neutrino Oscillation Experiment Using J-PARC
  Neutrino Beam and Hyper-Kamiokande},
\newblock 2014, 1412.4673.

\bibitem{Ghosh:2013zna}
Ghosh M, Ghoshal P, Goswami S, Raut SK,
\newblock Phys. Rev. D89:011301 (2014), 1306.2500.

\bibitem{Abe:2011ts}
Abe K, et~al.,
\newblock (2011), 1109.3262.

\bibitem{Razzaque:2014vba}
Razzaque S, Smirnov A{\relax Yu},
\newblock JHEP 05:139 (2015), 1406.1407.

\bibitem{Djurcic:2015vqa}
JUNO Collab., Djurcic Z, et~al.,
\newblock (2015), 1508.07166.

\bibitem{An:2015jdp}
JUNO, An F, et~al.,
\newblock J. Phys. G43:030401 (2016), 1507.05613.

\bibitem{Kim:2014rfa}
Kim SB,
\newblock Nucl. Part. Phys. Proc. 265:93 (2015), 1412.2199.

\bibitem{Choubey:2011zzq}
IDS-NF Consortium, Choubey S, et~al.,
\newblock (2011), 1112.2853.

\bibitem{Geer:1997iz}
Geer S,
\newblock Phys. Rev. D57:6989 (1998), hep-ph/9712290,
\newblock [Erratum: Phys. Rev.D59,039903(1999)].

\bibitem{Wildner:2014kia}
Wildner E, et~al.,
\newblock Phys. Rev. ST Accel. Beams 17:071002 (2014).

\bibitem{::2013kaa}
LAGUNA-LBNO Collab., Agarwalla SK, et~al.,
\newblock JHEP 05:094 (2014), 1312.6520.

\bibitem{Henderson:2008zz}
Henderson S, ed.,
\newblock {\em High-intensity, high-brightness hadron beams. Proceedings, 42nd
  Advanced Beam Dynamics Workshop, ABDW-HB'08, Nashville, USA, August 25-29,
  2008}, 2008.

\bibitem{Barger:2007yw}
Barger V, et~al.,
\newblock (2007), 0705.4396.

\bibitem{Rubbia:2010zz}
LAGUNA Consortium, Rubbia A,
\newblock Acta Phys. Polon. B41:1727 (2010).

\bibitem{Adams:2013qkq}
LBNE Collab., Adams C, et~al.,
\newblock (2013), 1307.7335.

\bibitem{Acciarri:2015uup}
DUNE, Acciarri R, et~al.,
\newblock (2015), 1512.06148.

\bibitem{Willis:1974gi}
Willis WJ, Radeka V,
\newblock Nucl. Instrum. Meth. 120:221 (1974).

\bibitem{Nygren:1976fe}
Nygren DR,
\newblock eConf C740805:58 (1974).

\bibitem{Chen:1976pp}
Chen HH, Condon PE, Barish BC, Sciulli FJ,
\newblock {A Neutrino detector sensitive to rare processes. I. A Study of
  neutrino electron reactions},
\newblock FERMILAB-PROPOSAL-0496, 1976.

\bibitem{lartpc}
Rubbia C,
\newblock The liquid-argon time projection chamber: A new concept for neutrino
  detector,
\newblock CERN-EP/77-08 (1977).

\bibitem{Mosel:2016ar}
Mosel U,
\newblock Ann. Rev. Nucl. Part. Sci. 66 (2016).

\bibitem{Proceedings:2015zqa}
Szczerbinska B, Worcester E, eds.,
\newblock {\em Proceedings, Workshop on Neutrino Interactions, Systematic
  uncertainties and near detector physics: Session of CETUP* 2014}, volume
  1680, 2015.

\bibitem{Abe:2015zbg}
Hyper-Kamiokande Collab., Abe K, et~al.,
\newblock PTEP 2015:053C02 (2015), 1502.05199.

\end{thebibliography}
\end{document}